\documentclass[twocolumn,preprintnumbers,amsmath,amssymb]{revtex4-1}

\usepackage[utf8]{inputenc} 
\usepackage{setspace}

\usepackage{amstext,amssymb,amsfonts}
\usepackage{bm}
\usepackage{amsmath}

\usepackage{graphicx}
\usepackage{graphics}
\usepackage{epsfig}
\usepackage{graphicx,latexsym,amsmath,amssymb,amsfonts}

\usepackage{txfonts}

\usepackage{hyperref}

\usepackage{amstext,amsmath,amssymb,amsfonts}
\usepackage{bm}

\usepackage{mathtools,slashed}

\usepackage{xcolor}
\usepackage{color}
\definecolor{lightblue}{rgb}{0.2,0.2,0.7}
\definecolor{darkblue}{rgb}{0,0.25,0.5}
\definecolor{redbrown}{rgb}{0.875,0.25,0.125}
\definecolor{darkgreen}{rgb}{0,0.5,0}

\newcommand{\bra}[1]{\ensuremath{\langle #1 \vert}}
\newcommand{\ket}[1]{\ensuremath{\vert #1  \rangle}}

\renewcommand{\b}[1]{\ensuremath{\mathbf{#1}}}

\renewcommand{\d}{\ensuremath{\text{d}}}
\newcommand{\Tr}{\ensuremath{\text{Tr}}}

\newcommand{\sr}{\ensuremath{\text{sr}}}

\DeclareMathOperator{\erf}{erf}

\newcommand{\ct}{\ensuremath{\tilde{c}}}
\newcommand{\mut}{\ensuremath{\tilde{\mu}}}
\DeclareMathOperator*{\minmax}{minmax}
\DeclareMathOperator*{\stat}{stat}

\begin{document}
\title{Relativistic short-range exchange energy functionals beyond the local-density approximation}
\author{Julien Paquier$^1$} \email{julien.paquier@lct.jussieu.fr}
\author{Emmanuel Giner$^1$}
\author{Julien Toulouse$^{1,2}$} \email{toulouse@lct.jussieu.fr}
\affiliation{$^1$Laboratoire de Chimie Th\'eorique (LCT), Sorbonne Universit\'e and CNRS, F-75005 Paris, France}
\affiliation{$^2$Institut Universitaire de France, F-75005 Paris, France}
\date{May 11, 2020}

\begin{abstract}
We develop relativistic short-range exchange energy functionals for four-component relativistic range-separated density-functional theory using a Dirac-Coulomb Hamiltonian in the no-pair approximation. We show how to improve the short-range local-density approximation exchange functional for large range-separation parameters by using the on-top exchange pair density as a new variable. We also develop a relativistic short-range generalized-gradient approximation exchange functional which further increases the accuracy for small range-separation parameters. Tests on the helium, beryllium, neon, and argon isoelectronic series up to high nuclear charges show that this latter functional gives exchange energies with a maximal relative percentage error of $3\%$. The development of this exchange functional represents a step forward for the application of four-component relativistic range-separated density-functional theory to chemical compounds with heavy elements.
\end{abstract}

\maketitle

\section{Introduction}

Range-separated density-functional theory (RS-DFT) (see, e.g., Refs.~\onlinecite{Sav-INC-96,TouColSav-PRA-04}) is an extension of Kohn-Sham density-functional theory (DFT)~\cite{KohSha-PR-65} which rigorously combines a wave-function method accounting for the long-range part of the electron-electron interaction with a complementary short-range density functional. RS-DFT has a faster basis convergence than standard wave-function theory (WFT)~\cite{FraMusLupTou-JCP-15} and can improve over usual Kohn-Sham density-functional approximations (DFAs) for the description of strong-correlation effects (see, e.g., Refs.~\onlinecite{HedTouJen-JCP-18,FerGinTou-JCP-19}) or weak intermolecular interactions (see, e.g., Refs.~\onlinecite{TayAngGalZhaGygHirSonRahLilPodBulHenScuTouPevTruSza-JCP-16,KalMusTou-JCP-19}). 

For the description of compounds with heavy elements, RS-DFT can be extended to a four-component relativistic framework~\cite{KulSau-CP-12,SheKneSau-PCCP-15,PaqTou-JCP-18}. In particular, in Refs.~\onlinecite{KulSau-CP-12,SheKneSau-PCCP-15}, second-order M{\o}ller-Plesset perturbation theory and coupled-cluster theory based on a no-pair Dirac-Coulomb Hamiltonian with long-range electron-electron interaction were combined with short-range non-relativistic exchange-correlation DFAs and applied to heavy rare-gas dimers. One limitation, at least in principle, in these works is the neglect of relativity in the short-range density functionals. It is thus desirable to develop appropriate short-range relativistic exchange-correlation DFAs for four-component RS-DFT in order to quantify the error due to the neglect of relativity in the short-range density functionals and possibly increase the accuracy of these approaches. As a first step toward this, in Ref.~\onlinecite{PaqTou-JCP-18} some of the present authors developed a short-range relativistic local density-functional approximation (srRLDA) exchange functional based on calculations on the relativistic homogeneous electron gas (RHEG) with the Coulomb and Coulomb-Breit electron-electron interactions in the no-pair approximation.

In the present work we test this srRLDA exchange functional on atomic systems, namely the helium, beryllium, neon, and argon isoelectronic series up to high nuclear charges $Z$, using a four-component Dirac-Coulomb Hamiltonian in the no-pair approximation. We reveal that, for these relativistic ions with large $Z$, the srRLDA exchange functional is quite inaccurate even for large values of the range-separation parameter $\mu$. We show how this functional can be improved by using the on-top exchange pair density as a new variable. Finally, we further improve the short-range relativistic exchange functional by constructing a generalized-gradient approximation (GGA), achieving a 3\% maximal relative energy error.

The paper is organized as follows. In Section~\ref{sec:rsdft} we lay out the formalism of RS-DFT for a four-component relativistic Dirac-Coulomb Hamiltonian in the no-pair approximation. In Section~\ref{sec:comput} we give the computational details for our calculations. In Section~\ref{sec:rsrlda} we test the srRLDA exchange functional and discuss its limitations. In Section~\ref{sec:ontop} we improve the srRLDA exchange functional by using the on-top exchange pair density. In Section~\ref{sec:pbe}, we construct and test short-range relativistic exchange GGAs. Finally, Section~\ref{sec:conclusions} contains our conclusions. In the Appendices, we derive the uniform coordinate scaling relation for the relativistic short-range exchange density functional and the expression of the on-top exchange pair-density in a four-component no-pair relativistic framework.

\section{Relativistic range-separated density-functional theory}
\label{sec:rsdft}

Let us first discuss the choice of the relativistic quantum many-particle theory on which to base relativistic RS-DFT and the general strategy that we follow in this work. Clearly, since RS-DFT combines WFT and DFT, we need a relativistic framework which is convenient for both of them. Relativistic Kohn-Sham DFT has been formulated based on quantum electrodynamics (QED)~\cite{RajCal-PRB-73,EngMulSpeDre-INC-95,Eng-INC-02}, even though the no-pair approximation~\cite{Suc-PRA-80,Mit-PRA-81} is normally introduced at a later stage for practical calculations. As regards WFT, the best tractable relativistic framework is the recently developed effective QED Hamiltonian~\cite{LiuLin-JCP-13,Liu-PR-14,Liu-IJQC-15,Liu-NSR-16,Liu-JCP-20} incorporating all QED effects obtained with non-retarded two-particle interactions. In the present work, we will stick however to the most common choice of the four-component Dirac-Coulomb Hamiltonian in the no-pair approximation, which can easily be used for both WFT and DFT. This relativistic framework can be derived in several ways (see, e.g., Refs.~\onlinecite{Kut-CP-12,LiuLin-JCP-13,Liu-PR-14,Liu-NSR-16,Liu-JCP-20}).

In the effective QED approach, the Hamiltonian is written in second quantization with normal ordering with respect to the vacuum state of empty positive-energy one-particle states and completely filled negative-energy one-particle states, while incorporating charge-conjugation symmetry. This Hamiltonian has a stable vacuum state and is physically meaningful. In this approach, the no-pair approximation, corresponding to projecting this Hamiltonian onto the many-electron subspace generated by positive-energy one-particle states, is just a convenient approximation (but in principle not necessary), akin to the idea of restricting the orbitals entering the wave function to an active orbital subspace in the complete-active-space self-consistent field method. The no-pair Dirac-Coulomb Hamiltonian is then obtained by further neglecting the effective QED one-particle potential corresponding to vacuum polarization and electron self-energy. By contrast, in the configuration-space approach, the Hamiltonian is written either in first quantization or, equivalently, in ``naive'' second quantization (i.e., without normal ordering with respect to a stable vacuum state). The resulting Hamiltonian has thus an unstable vacuum state, corresponding to empty positive-energy one-particle states and empty negative-energy one-particle states, it has no bound states (the electronic states that should be normally bound being embedded in the continuum of excitations to positive-energy states and deexcitations to negative-energy states), and hence is per se unphysical. However, by projecting this Hamiltonian onto the many-electron subspace generated by the positive-energy one-particle states, we recover the same physically relevant no-pair Dirac-Coulomb Hamiltonian as the one obtained by starting with the effective QED approach.

One drawback of the no-pair approximation is that the projector onto the subspace of electronic states depends on the separation between positive-energy and negative-energy one-particle states, and therefore depends on the potential used to generate these one-particle states. If the projector is applied to the Hamiltonian, the whole resulting projected Hamiltonian is thus dependent on this potential. As mentioned in Ref.~\onlinecite{Eng-INC-02}, this is problematic for formulating DFT since we cannot isolate, as normally done, an universal part of the Hamiltonian, and we thus cannot define universal density functionals. However, instead of thinking of the projector as being applied to the Hamiltonian, we can equivalently think of the projector as being applied to the considered many-electron state and optimize the projector simultaneously with the wave function. In this way, we can introduce universal density functionals, similarly to non-relativistic DFT, defined such that for a given density a constrained-search optimization of the projected wave function will determine alone the optimal projector without the need of pre-choosing a particular potential, at least for systems for which positive-energy and negative-energy one-particle states can be unambiguously separated. Again, both the effective QED approach or the configuration-space approach can a priori be used for doing so. In the effective QED approach, the projector would be optimized (by rotations between positive-energy and negative-energy one-particle states) using an energy minimization principle. In the configuration-space approach, the projector is optimized using a minmax principle~\cite{Tal-PRL-86,DatDev-PJP-88,GriSie-JLMS-99,DolEstSer-JFA-00,SauVis-INC-03,AlmKneJenDyaSau-JCP-16}. In the present work, we follow the configuration-space approach and leave for future work the alternative formulation based on the effective QED approach.

We thus start with the Dirac-Coulomb electronic Hamiltonian (see, e.g., Refs.~\onlinecite{DyaFae-BOOK-07,ReiWol-BOOK-09})
\begin{eqnarray}
\hat{H} &=& \hat{T}_{\text{D}} +  \hat{V}_{\text{ne}} + \hat{W}_{\text{ee}},
\end{eqnarray}
where the kinetic + rest mass Dirac operator $\hat{T}_{\text{D}}$, the nuclei-electron interaction operator $\hat{V}_{\text{ne}}$, and the Coulomb electron-electron interaction operator $\hat{W}_{\text{ee}}$ are expressed using four-component creation and annihilation field operators $\hat{\psi}^\dagger(\b{r})$ and $\hat{\psi}(\b{r})$ without normal reordering with respect to a stable vacuum state. We thus write $\hat{T}_{\text{D}}$ as
\begin{eqnarray}
\hat{T}_{\text{D}} &=&   \int \hat{\psi}^\dagger(\b{r}) \left[ c \; (\bm{\alpha} \cdot \b{p}) +\bm{\beta} \; mc^{2} \right] \hat{\psi}(\b{r}) \; \d\b{r}, \;\;\;\;\;
\label{TD}
\end{eqnarray}
where $\b{p} = -i \bm{\nabla}_{\b{r}}$ is the momentum operator, $c = 137.036$ a.u. is the speed of light, $m=1$ a.u. is the electron mass, and 
$\bm{\alpha}$ and $\bm{\beta}$ are the $4 \times 4$ Dirac matrices
\begin{eqnarray}
\bm{\alpha} = \left(\begin{array}{cc}
\b{0}_2&\bm{\sigma}\\
\bm{\sigma}&\b{0}_2\\
\end{array}\right)
~\text{and}~~
\bm{\beta} = \left(\begin{array}{cc}
\b{I}_{2}&\b{0}_2\\
\b{0}_2&-\b{I}_{2}\\
\end{array}\right),
\end{eqnarray}
where $\bm{\sigma}=(\bm{\sigma}_x,\bm{\sigma}_y,\bm{\sigma}_z)$ is the 3-dimensional vector of the $2 \times 2$ Pauli matrices, and $\b{0}_2$ and $\b{I}_2$ are the $2 \times 2$ zero and identity matrices, respectively. Similarly, we write $\hat{V}_{\text{ne}}$ and $\hat{W}_{\text{ee}}$ as
\begin{eqnarray}
\hat{V}_{\text{ne}} &=&   \int v_{\text{ne}}(\b{r}) \; \hat{n}(\b{r}) \; \d\b{r} , \;\;\;\;\; 
\label{Vne}
\end{eqnarray}
where $v_{\text{ne}}(\b{r})$ is the nuclei-electron potential, and
\begin{eqnarray}
\hat{W}_{\text{ee}} &=& \frac{1}{2} \iint w_{\text{ee}}(r_{12}) \; \hat{n}_2(\b{r}_1,\b{r}_2) \; \d\b{r}_1 \d\b{r}_2, \;\;\;\;
\label{Wee}
\end{eqnarray}
where $w_{\text{ee}}(r_{12})=1/r_{12}$ is the Coulomb electron-electron potential, and $\hat{n}(\b{r}) = \hat{\psi}^\dagger(\b{r}) \hat{\psi}(\b{r})$ and $\hat{n}_2(\b{r}_1,\b{r}_2) = \hat{\psi}^\dagger(\b{r}_1) \hat{\psi}^\dagger(\b{r}_2) \hat{\psi}(\b{r}_2) \hat{\psi}(\b{r}_1)$ are the density and pair density operators, respectively.

Introducing a set of orthonormal 4-component-spinor orbitals $\{\psi_p(\b{r})\}$ which are eigenfunctions of a one-particle Dirac Hamiltonian with some potential, and assuming that this set of orbitals can be partitioned into a set of positive-energy orbitals and a set of negative-energy orbitals, $\{\psi_p(\b{r})\}= \{\psi_p(\b{r})\}_{\varepsilon_p>0} \cup \{\psi_p(\b{r})\}_{\varepsilon_p<0}$, the no-pair~\cite{Suc-PRA-80,Mit-PRA-81} relativistic ground-state energy of a $N$-electron system can be defined using a minmax principle~\cite{Tal-PRL-86,DatDev-PJP-88,GriSie-JLMS-99,DolEstSer-JFA-00,SauVis-INC-03,AlmKneJenDyaSau-JCP-16}, that we will formally write as,
\begin{eqnarray}
E_0  &=& \minmax_{\Psi_+} \; \bra{\Psi_+} \hat{T}_\text{D} + \hat{V}_\text{ne} + \hat{W}_{\text{ee}} \ket{\Psi_+}.
\label{EWFT}
\end{eqnarray}
In this equation, we search over normalized wave functions of the form $\ket{\Psi_+} = \hat{P}_+ \ket{\Psi}$, where $\hat{P}_+$ is the projector on the $N$-electron-state space generated by the set of positive-energy orbitals $\{\psi_p(\b{r})\}_{\varepsilon_p>0}$ and $\ket{\Psi}$ is a general $N$-electron antisymmetric wave function, and the notation ${\minmax_{\Psi_+}= \min_{\Psi} \max_{\hat{P}_+} = \max_{\hat{P}_+} \min_{\Psi}}$ means a minimization with respect to $\Psi$ and a maximization with respect to $\hat{P}_+$. This maximization must be done by rotations of the positive-energy orbitals $\{\psi_p(\b{r})\}_{\varepsilon_p>0}$ with its complement set of negative-energy orbitals $\{\psi_p(\b{r})\}_{\varepsilon_p<0}$. Here, we have assumed that the optimum of the minmax is a saddle point in the wave-function parameter space (which can be calculated with a multiconfiguration self-consistent-field (MCSCF) algorithm~\cite{JenDyaSauFae-JCP-96,ThyFleJen-JCP-08,AlmKneJenDyaSau-JCP-16}), so that the same energy is obtained whatever the order of $\min_{\Psi}$ and $\max_{\hat{P}_+}$. Note that, in the non-relativistic limit ($c\to\infty$), the energy gap between positive- and negative-energy orbitals of order $2mc^2$ goes to infinity and the maximization over $\hat{P}_+$ becomes useless, and thus the minmax principle properly reduces to the non-relativistic minimization principle.

Now we attempt to formulate a relativistic DFT within this no-pair approximation. Following the spirit of the constrained-search formulation of non-relativistic DFT~\cite{Lev-PNAS-79,Lie-IJQC-83}, we propose to define the no-pair relativistic universal density functional as
\begin{eqnarray}
F[n]  &=& \minmax_{\Psi_+\to n} \; \bra{\Psi_+} \hat{T}_\text{D} + \hat{W}_{\text{ee}} \ket{\Psi_+}
\nonumber\\
      &=& \bra{\Psi_+[n]} \hat{T}_\text{D} + \hat{W}_{\text{ee}} \ket{\Psi_+[n]},
\label{F}
\end{eqnarray}
where the minmax procedure is identical to that in Eq.~(\ref{EWFT}) except for the additional constraint that $\Psi_+$ yields the density $n$, i.e. $\bra{\Psi_+} \hat{n}(\b{r}) \ket{\Psi_+} = n(\b{r})$. In Eq.~(\ref{F}), $\Psi_+[n]$ is the optimal wave function for the density $n$. We will again assume that the optimum of the minmax is a saddle point in the density-constrained wave-function parameter subspace. Of course, this functional is only defined for densities which come from a wave function of the form of $\Psi_+$, which we will refer to as $\Psi_+$-representable densities. Note that, consistently with neglecting the Breit electron-electron interaction, we will only consider functionals of the density and not of the density current. The no-pair relativistic ground-state energy of Eq.~(\ref{EWFT}) can be in principle obtained from $F[n]$ as a stationary point with respect to variations over $\Psi_+$-representable densities
\begin{eqnarray}
E_0 \in  \stat_{n} \left\{ F[n] + \int v_\text{ne}(\b{r}) \; n(\b{r}) \; \d\b{r} \right\},
\label{Estat}
\end{eqnarray}
where we have introduced the notation $\stat_n$ to designates the set of stationary energies with respect to variations of $n$. Due to the minmax principle in Eqs.~(\ref{EWFT}) and (\ref{F}), we can only assume a stationary principle in Eq.~(\ref{Estat}), instead of the usual non-relativistic minimization principle over densities. This situation is in fact similar to the problem of formulating a pure-state time-independent variational extension of DFT for excited-state energies~\cite{Gor-PRA-99,AyeLev-PRA-09}. 

We now define a no-pair relativistic long-range universal density functional, similarly to Eq.~(\ref{F}), as
\begin{eqnarray}
F^{\text{lr},\;\mu}[n]  &=& \minmax_{\Psi_+\to n} \; \bra{\Psi_+} \hat{T}_\text{D} + \hat{W}_{\text{ee}}^{\text{lr},\;\mu} \ket{\Psi_+}
\nonumber\\
&=& \bra{\Psi_+^\mu[n]} \hat{T}_\text{D} + \hat{W}_{\text{ee}}^{\text{lr},\;\mu} \ket{\Psi_+^\mu[n]},
\label{Flr}
\end{eqnarray}
with the long-range electron-electron interaction operator $\hat{W}_{\text{ee}}^{\text{lr},\;\mu} = (1/2) \iint w_{\text{ee}}^{\text{lr},\;\mu}(r_{12}) \; \hat{n}_2(\b{r}_1,\b{r}_2) \; \d\b{r}_1 \d\b{r}_2$ where $w_{\text{ee}}^{\text{lr},\;\mu}(r_{12})=\erf(\mu r_{12})/r_{12}$ is the long-range electron-electron potential and $\mu$ is the range-separation parameter. In Eq.~(\ref{Flr}) $\Psi_+^\mu[n]$ is the optimal wave function for the density $n$ and range-separation parameter $\mu$. We can thus decompose the density functional $F[n]$ as
\begin{eqnarray}
F[n]  &=& F^{\text{lr},\;\mu}[n] + \bar{E}_\text{Hxc}^{\text{sr},\;\mu}[n],
\label{Fdecomp}
\end{eqnarray}
which defines the complement relativistic short-range Hartree-exchange-correlation density functional $\bar{E}_\text{Hxc}^{\text{sr},\;\mu}[n]$. Plugging Eq.~(\ref{Fdecomp}) into Eq.~(\ref{Estat}), we conclude that the no-pair relativistic ground-state energy of Eq.~(\ref{EWFT}) corresponds to a stationary point of the following range-separated energy expression over $\Psi_+$ wave functions
\begin{eqnarray}
E_0 \in  \stat_{\Psi_+} \; \left\{ \bra{\Psi_+} \hat{T}_\text{D} + \hat{V}_\text{ne} + \hat{W}_{\text{ee}}^{\text{lr},\;\mu} \ket{\Psi_+} + \bar{E}_\text{Hxc}^{\text{sr},\;\mu}[n_{\Psi_+}] \right\},
\label{EstatPsi}
\end{eqnarray}
where $n_{\Psi_+}$ is the density of $\Psi_+$. For practical calculations, we will \textit{assume} that the no-pair relativistic ground-state energy corresponds in fact to the minmax search over $\Psi_+$
\begin{eqnarray}
E_0  &=& \minmax_{\Psi_+} \; \left\{ \bra{\Psi_+} \hat{T}_\text{D} + \hat{V}_\text{ne} + \hat{W}_{\text{ee}}^{\text{lr},\;\mu} \ket{\Psi_+} + \bar{E}_\text{Hxc}^{\text{sr},\;\mu}[n_{\Psi_+}] \right\}.\;\;\;\;\;
\label{EminmaxRSDFT}
\end{eqnarray}
Even though we do not see any guarantee that this is always true, it seems a reasonable working assumption for practical calculations. In fact, it corresponds to what is done in practice in no-pair Kohn-Sham DFT calculations~\cite{Raj-JPC-78,MacVos-JPC-79,EngKelFacMulDre-PRA-95,LiuHonDaiLiDol-TCA-97,VarFriNakMukAntGesHeiEngBas-JCP-00,YanIikNakIshHir-JCP-01,SauHel-JCC-02,QuiBel-JCP-02,KomRepMalMalOnKau-JCP-08,BelStoQuiTar-PCCP-11}, which corresponds to Eq.~(\ref{EminmaxRSDFT}) in the special case of $\mu=0$, i.e. 
\begin{eqnarray}
E_0  &=& \minmax_{\Phi_+} \; \left\{ \bra{\Phi_+} \hat{T}_\text{D} + \hat{V}_\text{ne} \ket{\Phi_+} + E_\text{Hxc}[n_{\Phi_+}] \right\},
\label{EminmaxKSDFT}
\end{eqnarray}
where the wave function can be restricted to a single determinant $\Phi_+$ and $E_\text{Hxc}[n]$ is the relativistic Kohn-Sham Hartree-exchange-correlation density functional. Another special case of Eq.~(\ref{EminmaxRSDFT}) is for $\mu\to\infty$ for which we correctly recover the wave-function theory of Eq.~(\ref{EWFT}).

As usual, we can decompose the complement relativistic short-range Hartree-exchange-correlation density functional into separate components
\begin{eqnarray}
\bar{E}_\text{Hxc}^{\text{sr},\;\mu}[n] = E_\text{H}^{\text{sr},\;\mu}[n] + E_\text{x}^{\text{sr},\;\mu}[n] + \bar{E}_\text{c}^{\text{sr},\;\mu}[n].
\end{eqnarray}
In this expression, $E_{\text{H}}^{\text{sr},{\mu}}[n]$ is the short-range Hartree density functional (which has the same expression as in the non-relativistic case)
\begin{eqnarray}
E_{\text{H}}^{\text{sr},{\mu}}[n] = \frac{1}{2} \iint w_{\text{ee}}^{\text{sr},\;\mu}(r_{12}) \; n(\b{r}_1)n(\b{r}_2) \; \d\b{r}_1 \d\b{r}_2,
\end{eqnarray}
where $w_{\text{ee}}^{\text{sr},\;\mu}(r_{12}) = w_{\text{ee}}(r_{12}) - w_{\text{ee}}^{\text{lr},\;\mu}(r_{12})$ is the short-range electron-electron potential, $E_\text{x}^{\text{sr},\;\mu}[n]$ is the relativistic short-range exchange density functional
\begin{eqnarray}
E_{\text{x}}^{\text{sr},{\mu}}[n] = \bra{\Phi_+[n]} \; \hat{W}_\text{ee}^{\text{sr},{\mu}} \; \ket{\Phi_+[n]} - E_{\text{H}}^{\text{sr},{\mu}}[n],
\label{Exsrmu}
\end{eqnarray}
where $\Phi_+[n] = \Psi_+^{\mu=0}[n]$ is the relativistic Kohn-Sham single-determinant wave function and $\hat{W}_{\text{ee}}^{\text{sr},\;\mu} = (1/2) \iint w_{\text{ee}}^{\text{sr},\;\mu}(r_{12}) \; \hat{n}_2(\b{r}_1,\b{r}_2) \; \d\b{r}_1 \d\b{r}_2$ is the short-range electron-electron interaction operator, and $\bar{E}_\text{c}^{\text{sr},\;\mu}[n]$ is the complement relativistic short-range correlation density functional. In Appendix~\ref{app:scaling}, we show that the relativistic short-range exchange density functional $E_{\text{x}}^{\text{sr},{\mu}}[n]$ satisfies a uniform coordinate scaling relation [Eq.~(\ref{Exsrgamma})] which represents an important constraint to impose in approximations. 

Even though the present formulation of relativistic RS-DFT seems reasonable for practical chemical applications, it obviously calls for a closer mathematical examination of its domain of validity. In particular, it is clear that the minmax principle of the no-pair approximation in the configuration-space approach breaks down in the strong relativistic regime (i.e., for nuclear charges $Z \gtrsim c$). Of course, in the non-relativistic limit ($c\to\infty$), relativistic RS-DFT properly reduces to non-relativistic RS-DFT.

\section{Computational setup}
\label{sec:comput}

We consider the helium, beryllium, neon, and argon isoelectronic series, up to the uranium nuclear charge $Z=92$. The electronic density $n(\b{r})$ naturally increases at the nucleus with $Z$ and can be conveniently measured with $k_{\text{F}_{\text{max}}}$, i.e. the maximal value taken at the nucleus by the local Fermi wave vector $k_{\text{F}}(\b{r}) = (3{\pi}^{2}n(\b{r}))^{1/3}$. The strength of the relativistic effects can be measured by comparing the local Fermi wave vector $k_{\text{F}}(\b{r})$ to the speed of light $c \simeq$ 137.036 a.u. (with $\hbar=m_{\text{e}}=$ 1 a.u.): very little relativistic effects are expected in regions where $k_{\text{F}}(\b{r}) \ll c$, while strong relativistic effects are expected in regions where $k_{\text{F}}(\b{r}) \gtrsim c$.

To test the different functionals, we have first performed four-component Dirac Hartree-Fock (DHF) calculations based on the relativistic Dirac-Coulomb Hamiltonian with point-charge nucleus, using our own program implemented as a plugin of the software {\sc Quantum package 2.0}~\cite{QP-JCTC-19}. For the helium series, we use the dyall\_1s2.3z basis set of Ref.~\onlinecite{AlmKneJenDyaSau-JCP-16} except for $\text{Yb}^{68+}$ and $\text{U}^{90+}$ for which the basis set was not available. For these systems, as well as for the beryllium, neon, and argon series, we construct uncontracted even-tempered Gaussian-type orbital basis sets~\cite{Fae-SD-04}, following the primitive structure of the dyall-cvdz basis sets for He, Be, Ne, and Ar~\cite{Dya-TCA-16}. For each system and angular momentum, the exponents of the large-component basis functions are taken as the geometric series
\begin{eqnarray}
{\zeta}_{\nu} &=& {\zeta}_{1}~q^{{\nu}-1},
\end{eqnarray}
where ${\zeta}_{1}$ is chosen among the largest exponents from the dyall-cvdz basis set for the given element and angular momentum~\cite{Dya-TCA-16,Dya-TCA-06}, and the parameter $q$ is optimized by minimizing the DHF total energy. The small-component basis functions are generated from the unrestricted kinetic-balance scheme~\cite{DyaFae-Chem.Phys.Letters-90}. The basis-set parameters are given in the Supplementary Material.

Using the previously obtained DHF orbitals, we then estimate the short-range exact exchange energy
\begin{equation}
E_\text{x}^{\text{sr},\mu} = \frac{1}{2} \iint w_\text{ee}^{\sr,\mu}(r_{12}) \; n_{2,\text{x}}(\b{r}_1,\b{r}_2) \d\b{r}_1 \d\b{r}_2,
\end{equation}
where $n_{2,x}(\b{r}_1,\b{r}_2)$ is the exchange pair density
\begin{equation}
n_{2,\text{x}}(\b{r}_1,\b{r}_2) = - \Tr[ \gamma(\b{r}_1,\b{r}_2) \gamma(\b{r}_2,\b{r}_1)],
\end{equation}
and $\gamma(\b{r}_1,\b{r}_2) = \sum_{i=1}^N \psi_i(\b{r}_1) \psi_i^{\dagger}(\b{r}_2)$ is the $4\times4$ one-electron density matrix written with the four-component spinor occupied orbitals $\{ \psi_i(\b{r}) \}$. This short-range DHF exchange energy is used as the reference for testing the different exchange energy functionals, which are evaluated with the DHF density $n(\b{r}) = \Tr[ \gamma(\b{r},\b{r})]$ (and the DHF exchange on-top pair density for some of them, see below) using a SG-2-type quadrature grid~\cite{DasHer-JCC-17} with the radial grid of Ref.~\onlinecite{MurKno-JCP-96}.

\section{Short-range exchange local-density approximations}
\label{sec:rsrlda}

The non-relativistic short-range local-density approximation (srLDA) for the exchange functional has the expression
\begin{equation}
E_\text{x}^{\text{sr,LDA},\,\mu}[n] = \int n(\b{r}) \, \epsilon_\text{x}^{\text{sr,HEG},\,\mu}(n(\b{r})) \; \d\b{r},
\end{equation}
where the non-relativistic short-range homogeneous electron gas (HEG) exchange energy per particle $\epsilon_\text{x}^{\text{sr,HEG},\,\mu}(n)$ can be found in Refs.~\onlinecite{GilAdaPop-MP-96,Sav-INC-96,TouSavFla-IJQC-04}. The relativistic generalization of this functional, referred to as srRLDA, is
\begin{equation}
E_\text{x}^{\text{sr,RLDA},\,\mu}[n] = \int n(\b{r}) \, \epsilon_\text{x}^{\text{sr,RHEG},\,\mu}(n(\b{r})) \; \d\b{r},
\end{equation}
where the short-range RHEG exchange energy per particle $\epsilon_\text{x}^{\text{sr,RHEG},\,\mu}(n)$ is given in Ref.~\onlinecite{PaqTou-JCP-18} with arbitrary accuracy as systematic Padé approximants with respect to the dimensionless variable $\ct = c/k_{\text{F}}= c/(3\pi^2 n)^{1/3}$ (we employ here the Padé approximant of order 6) with coefficients written as functions of the dimensionless range-separation parameter $\mu/k_\text{F}$. The dependence of $\epsilon_\text{x}^{\text{sr,RHEG},\,\mu}(n)$ on the dimensionless parameters $\ct$ and $\mu/k_\text{F}$ is a consequence of the uniform coordinate scaling relation of Eq.~(\ref{Exsrgamma}) which is valid of the RHEG.

\begin{figure}[t]
        \centering
        \includegraphics[width=5cm,angle=270]{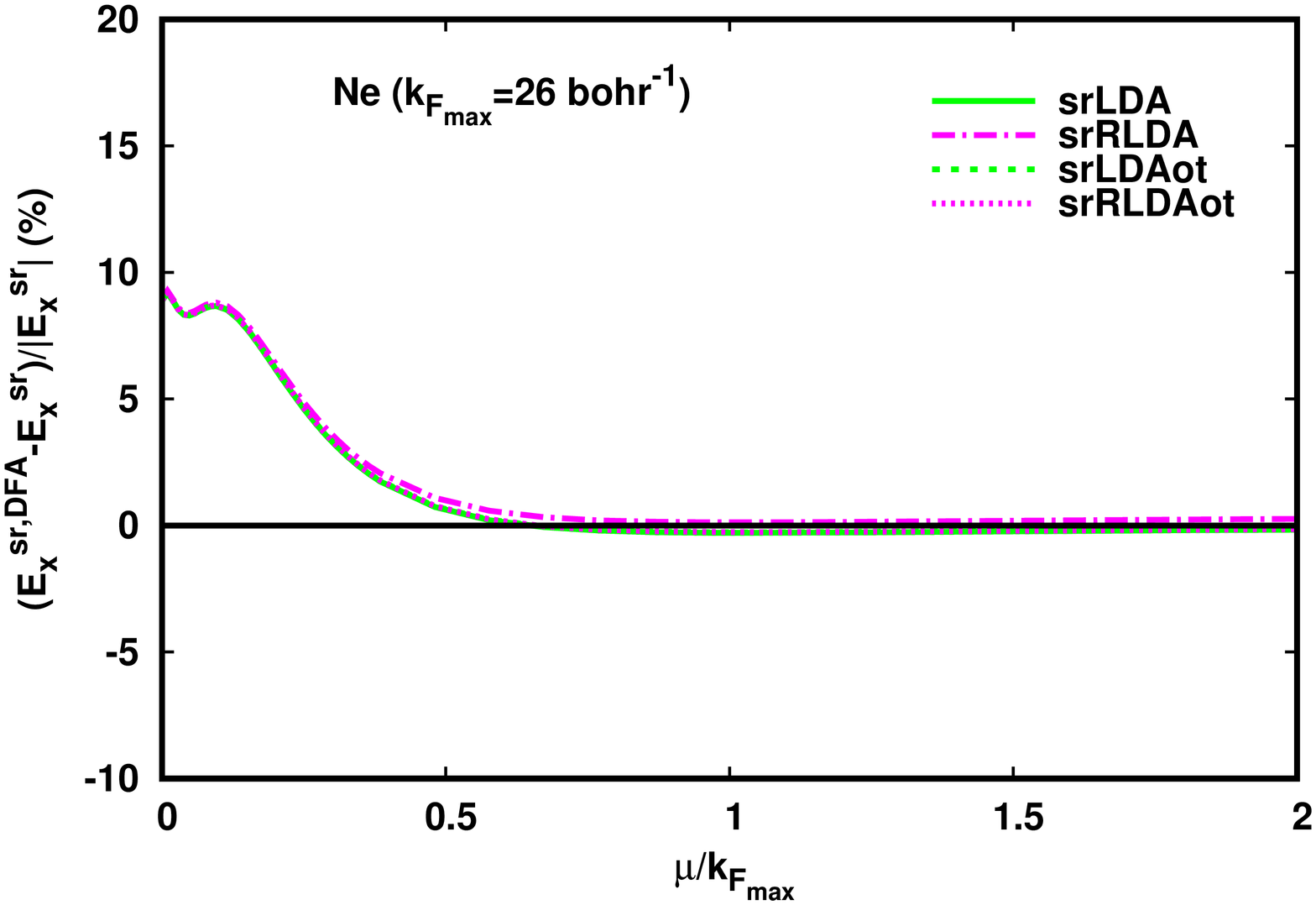}
        \includegraphics[width=5cm,angle=270]{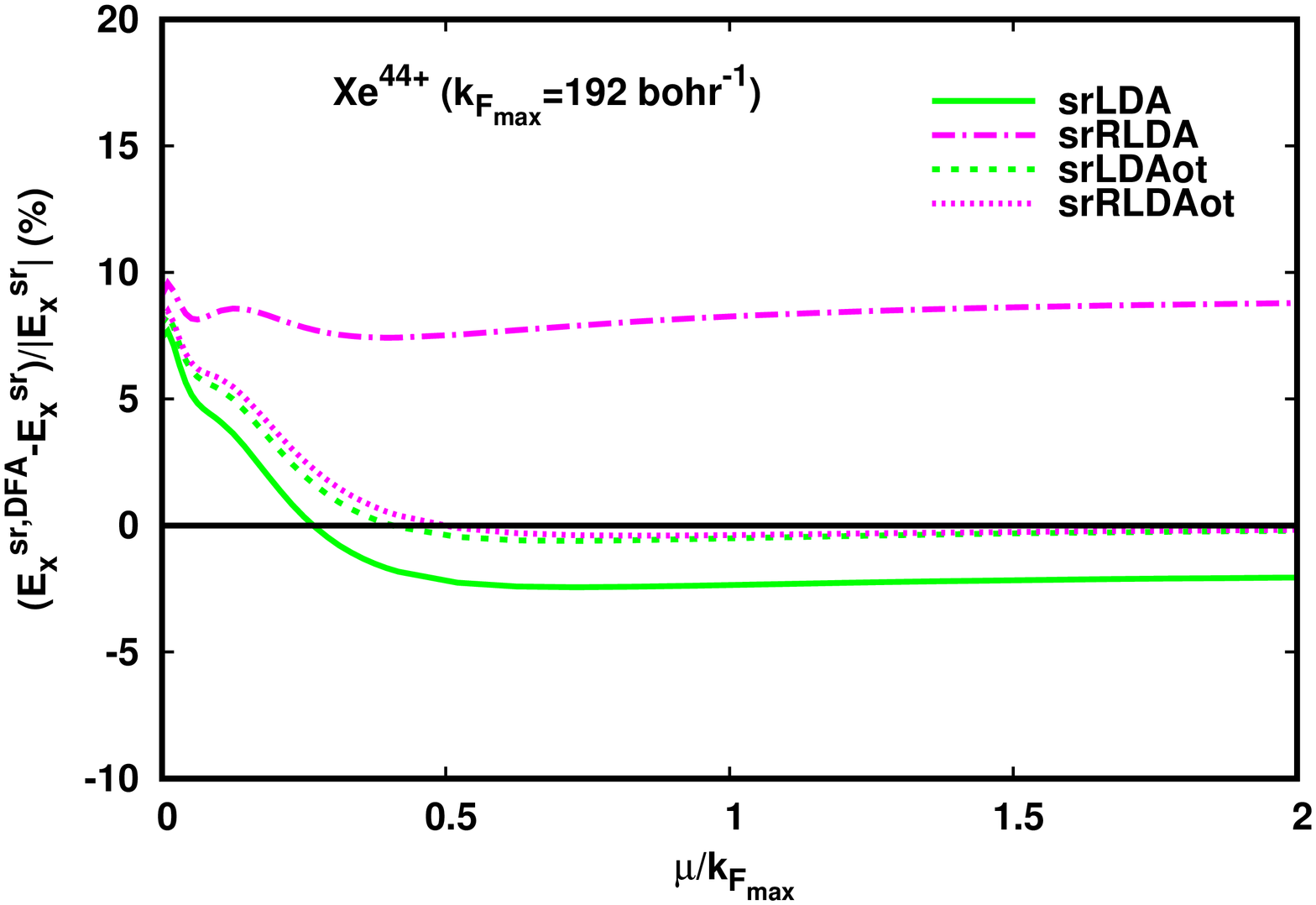}
        \includegraphics[width=5cm,angle=270]{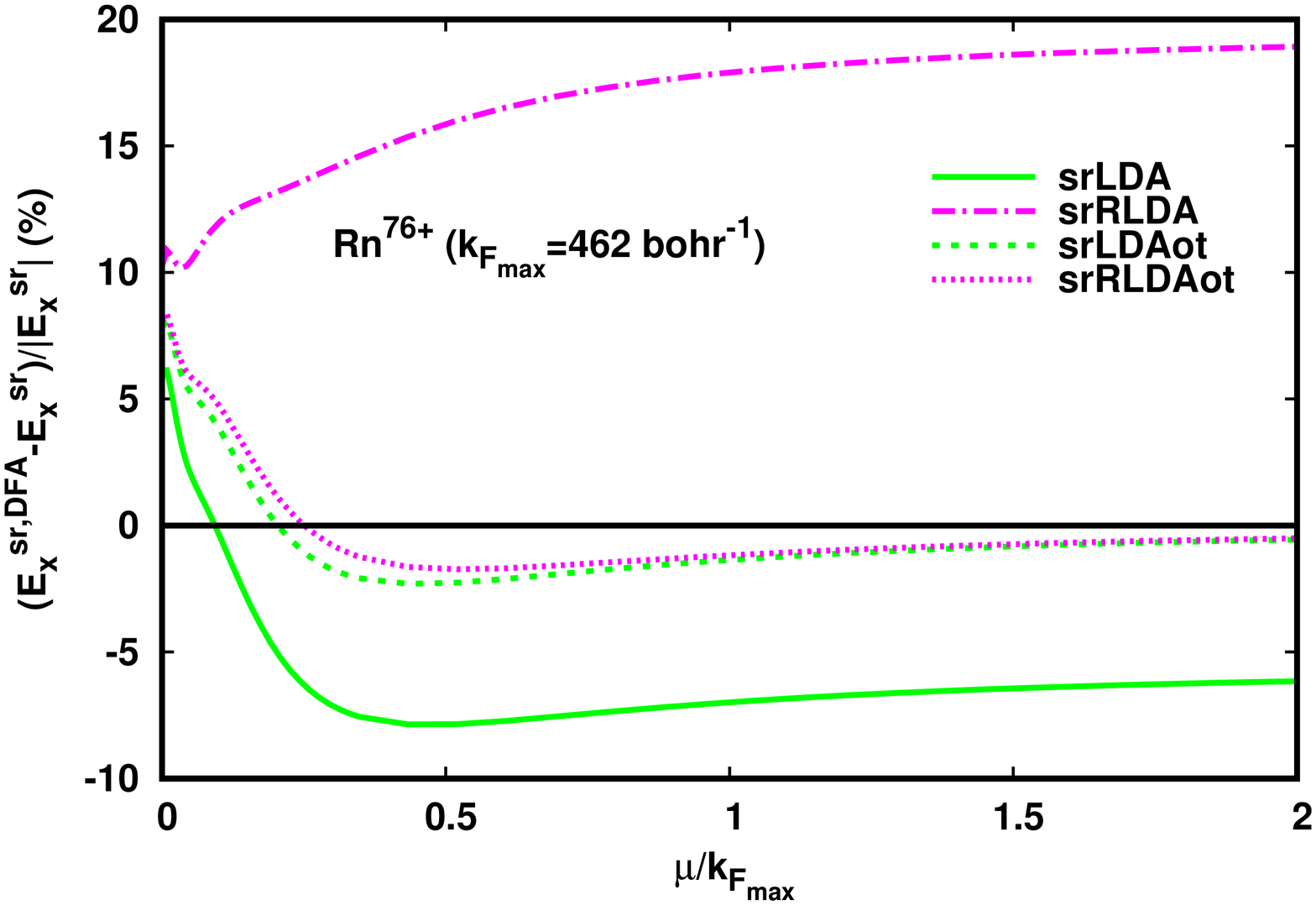}
\caption{Relative percentage error on the short-range exchange energy calculated with the srLDA, srRLDA, srLDAot, srRLDAot functionals for three representative members of the neon isoelectronic series (Ne, Xe$^{44+}$, and Rn$^{76+}$).}
\label{fig:RLDA}
\end{figure}

The relative percentage errors of the srLDA and srRLDA exchange functionals with respect to the short-range DHF exchange energy, i.e. $100 \times(E_{\text{x}}^{\text{sr,DFA},\,\mu} - E_{\text{x}}^{\text{sr},\,\mu})/|E_{\text{x}}^{\text{sr},\,\mu}|$, are plotted in Fig.~\ref{fig:RLDA} as a function of the dimensionless range-separation parameter ${\mu}/k_{\text{F}_{\text{max}}}$ for three representative members of the neon isoelectronic series (Ne, Xe$^{44+}$ and Rn$^{76+}$). The relativistic effects go from very small for Ne to very large for Rn$^{76+}$.

For $\mu=0$, the short-range interaction reduces to the full-range Coulomb interaction, and we observe that both the non-relativistic and relativistic LDA exchange functionals underestimate (in absolute value) the DHF exchange energy by 5\% to 10 \%. As previously noted~\cite{EngKelDre-PRA-96}, the non-relativistic LDA exchange functional (evaluated with a relativistic density) fortuitously gives exchange energies with lower errors than the relativistic LDA exchange functional for systems with significant relativistic effects (Xe$^{44+}$, and Rn$^{76+}$). When $\mu$ increases, the srLDA and srRLDA exchange functionals show quite different behaviors for these relativistic systems. The relative error of the srLDA exchange energy changes sign with $\mu$ and eventually goes to a negative constant for $\mu \to \infty$, corresponding to an overestimation in absolute value. By contrast, the relative error of the srRLDA exchange energy always remains positive and goes to a positive constant for $\mu \to \infty$, corresponding to an underestimation in absolute value. The more relativistic the system is, the largest this overestimation or underestimation is. While for Ne at large $\mu$ both the srLDA and srRLDA exchange functionals have almost vanishing relative errors, for Rn$^{76+}$ at large $\mu$ the srLDA exchange energy is too negative by a little more than 5\% and the srRLDA exchange energy is too positive by almost 20\%. Clearly, both the srLDA and srRLDA exchange functionals are not accurate for relativistic systems.

In non-relativistic theory, it is known that the srLDA exchange functional becomes exact for large $\mu$~\cite{TouColSav-PRA-04}, which is one of the key advantages of RS-DFT. As apparent from Figure~\ref{fig:RLDA}, for relativistic systems, this nice property does not hold anymore for both the srLDA and srRLDA exchange functionals. This observation can be understood by using the distributional asymptotic expansion of the short-range interaction for large $\mu$~\cite{TouColSav-PRA-04}
\begin{eqnarray}
w^{\text{sr},\,{\mu}}_{\text{ee}}(r_{12}) &=&   \frac{{\pi}}{{\mu}^2} {\delta}(\b{r}_{12}) + \large{O}\left(\frac{1}{{\mu}^3}\right),
\end{eqnarray}
which directly leads to the asymptotic expansion of the short-range exact exchange energy
\begin{eqnarray}
E_{\text{x}}^{\text{sr},{\mu}} &=& \frac{{\pi}}{2~{\mu}^2} \int n_{2,\text{x}}(\b{r},\b{r})~\d\b{r} + \large{O} \left(\frac{1}{{\mu}^{4}}\right),
\label{Exsrmuasymp}
\end{eqnarray}
where $\displaystyle{n_{2,\text{x}}(\b{r},\b{r})}$ is the on-top exchange pair density. In the non-relativistic theory, considering the case of closed-shell systems for the sake of simplicity, the on-top exchange pair density is simply given in terms of the density as~\cite{ZieRauBae-TCA-77}
\begin{eqnarray}
n_{2,\text{x}}^{\text{NR}}(\b{r},\b{r}) &=& -\frac{n(\b{r})^{2}}{2}, 
\end{eqnarray}
and the srLDA exchange functional becomes indeed exact for large $\mu$
\begin{eqnarray}
E_{\text{x}}^{\text{sr,LDA},{\mu}}[n] &=& \frac{{\pi}}{2~{\mu}^2} \int n_{2,\text{x}}^{\text{HEG},0}(n(\b{r}))~\d\b{r} + \large{O} \left(\frac{1}{{\mu}^{4}}\right),
\label{ExsrLDAlargemu}
\end{eqnarray}
with the on-top exchange pair density of the non-relativistic HEG
\begin{eqnarray}
n_{2,\text{x}}^{\text{HEG},0}(n)=-\frac{n^2}{2}.
\end{eqnarray}
In the relativistic theory, the on-top exchange pair density is no longer a simple function of the density
\begin{eqnarray}
n_{2,\text{x}}(\b{r},\b{r}) = - \Tr[ \gamma(\b{r},\b{r})^2],
\end{eqnarray}
which is not equal to $-n(\b{r})^{2}/2$, except in the special case of two electrons in a unique Kramers pair (see Appendix~\ref{app:ontoprelat}). Therefore, for relativistic systems with more than two electrons, we see that the srLDA exchange functional is not exact for large $\mu$ [Eq.~(\ref{ExsrLDAlargemu})]. The srRLDA exchange functional is also not exact for large $\mu$. It takes the form
\begin{eqnarray}
E_{\text{x}}^{\text{sr,RLDA},{\mu}}[n] &=& \frac{{\pi}}{2~{\mu}^2} \int n_{2,\text{x}}^{\text{RHEG},0}(n(\b{r}))~\d\b{r} + \large{O} \left(\frac{1}{{\mu}^{4}}\right),
\label{ExsrRLDAlargemu}
\end{eqnarray}
with the on-top exchange pair density of the RHEG
\begin{eqnarray}
n_{2,\text{x}}^{\text{RHEG},0}(n)=-\frac{n^2}{4} (1 + h(\ct)),
\end{eqnarray}
and the function~\cite{PaqTou-JCP-18}
\begin{eqnarray}
        h(\ct) &=& \frac{9}{4} \Bigg[ {\ct}^{2} + {\ct}^{4} 
\nonumber\\
&&- {\ct}^{4} \text{arcsinh}\left(\frac{1}{\ct}\right) \Bigg(2\sqrt{1+{\ct}^{2}} - {\ct}^{2}~\text{arcsinh}\left(\frac{1}{\ct}\right)\Bigg) \Bigg].
\end{eqnarray}
For an alternative but equivalent expression for $n_{2,\text{x}}^{\text{RHEG},0}(n)$, see Eq. (A1) of Ref.~\onlinecite{EngKelFacMulDre-PRA-95}. The srRLDA exchange functional is in fact not even exact at large $\mu$ for two electrons in a unique Kramers pair. In Section~\ref{sec:ontop}, we show how to impose the large-$\mu$ behavior on the srLDA and srRLDA exchange functionals.

\section{Short-range exchange local-density approximations with on-top exchange pair density}
\label{sec:ontop}

In order to impose the correct large-${\mu}$ behavior of the srLDA and srRLDA exchange functionals for relativistic systems, we need to introduce a new ingredient in these functionals, namely the exact (relativistic) on-top exchange pair density $n_{2,\text{x}}(\b{r},\b{r})$, or equivalently the on-top exchange hole
\begin{eqnarray}
n_{\text{x}}(\b{r},\b{r}) &=& \frac{n_{2,\text{x}}(\b{r},\b{r})}{n(\b{r})}. 
\end{eqnarray} 

A simple way to use $n_{\text{x}}(\b{r},\b{r})$ to correct the srLDA exchange functional is to find, at each position $\b{r}$, the effective density $n_\text{eff}(\b{r})$ at which the on-top exchange hole of the HEG, $n_{\text{x}}^{\text{HEG},0}(n) = n_{2,\text{x}}^{\text{HEG},0}(n)/n = -n/2$, is equal to the on-top exchange hole of the inhomogeneous system considered, $n_{\text{x}}(\b{r},\b{r})$, i.e.
\begin{eqnarray}
n_{\text{x}}^{\text{HEG},0}(n_\text{eff}(\b{r})) = n_{\text{x}}(\b{r},\b{r}),
\label{nxHEG0neff}
\end{eqnarray}
which simply gives $n_\text{eff}(\b{r}) = -2 n_{\text{x}}(\b{r},\b{r})$. We then define the srLDA exchange functional with the on-top exchange pair density (srLDAot) using this effective density as
\begin{eqnarray}
E_{\text{x}}^{\text{sr,LDAot},\,{\mu}} [n]= \int n(\b{r})~{\epsilon}_{\text{x}}^{\text{sr,HEG},\,{\mu}}(n_{\text{eff}}(\b{r})) ~\d\b{r}.
\end{eqnarray}
This approximation could be considered either as an implicit functional of the density alone since $n_{\text{x}}(\b{r},\b{r})$ is an implicit functional of the density through the orbitals, or as an explicit functional of both the density and the on-top exchange hole $n_{\text{x}}(\b{r},\b{r})$.
This approximation corresponds to changing the transferability criterion in the LDA: at a given point $\b{r}$, instead of taking the exchange energy per particle of the HEG having the same density than the inhomogeneous system at that point, we now take the exchange energy per particle of the HEG having the same on-top exchange hole than the inhomogeneous system at that point. Interestingly, this approximation can be thought of as a particular application of the recently formalized connector theory~\cite{Van-THESIS-18,VanAouPanGatRei-arXiv-19}.

Similarly, we can correct the srRLDA exchange functional by finding, at each position $\b{r}$, the effective density $n_\text{eff}^\text{R}(\b{r})$ at which the on-top exchange hole of the RHEG, $n_{\text{x}}^{\text{RHEG},0}(n) = n_{2,\text{x}}^{\text{RHEG},0}(n)/n = -(n/4)(1+h(\ct))$, is equal to the on-top exchange hole of the inhomogeneous system considered, $n_{\text{x}}(\b{r},\b{r})$, i.e.
\begin{eqnarray}
n_{\text{x}}^{\text{RHEG},0}(n_\text{eff}^\text{R}(\b{r})) = n_{\text{x}}(\b{r},\b{r}).
\label{nxRHEG0neff}
\end{eqnarray}
This equation is less trivial to solve than Eq.~(\ref{nxHEG0neff}) since $n_{\text{x}}^{\text{RHEG},0}(n)$ is a complicated nonlinear function of $n$ (through $\ct$). However, at each point $\b{r}$, a unique solution $n_\text{eff}^\text{R}(\b{r})$ exists since the function $n \mapsto n_{\text{x}}^{\text{RHEG},0}(n)$ is monotonically decreasing and spans the domain $]-\infty,0]$ in which $n_{\text{x}}(\b{r},\b{r})$ necessarily belongs. In practice, we easily find $n_\text{eff}^\text{R}(\b{r})$ by a numerical iterative method, and we use it to define the srRLDA exchange functional with the on-top exchange pair density (srRLDAot) as
\begin{eqnarray}
E_{\text{x}}^{\text{sr,RLDAot},\,{\mu}} [n]= \int n(\b{r})~{\epsilon}_{\text{x}}^{\text{sr,RHEG},\,{\mu}}(n_\text{eff}^\text{R}(\b{r})) ~\d\b{r}.
\end{eqnarray}
Both the srLDAot and srRLDAot exchange functionals now fulfill the exact asymptotic expansion for large $\mu$ [Eq.~(\ref{Exsrmuasymp})]. In fact, restoring the correct on-top value of the exchange hole could be beneficial for any value of $\mu$, given the fact that the accuracy of non-relativistic Kohn-Sham exchange DFAs has been justified by the exactness of the underlying LDA on-top exchange hole (in addition to fulfilling the correct sum rule of the exchange hole)~\cite{BurPerErn-JCP-98}. Finally, we note that, in the non-relativistic limit ($c\to\infty$), we have $n_\text{eff}(\b{r}) = n_\text{eff}^\text{R}(\b{r}) = n(\b{r})$ and all these short-range exchange functionals reduce to the non-relativistic srLDA exchange functional (i.e., srLDAot = srRLDAot = srRLDA = srLDA).

The relative percentage errors of the srLDAot and srRLDAot exchange functionals for Ne, Xe$^{44+}$, and Rn$^{76+}$ are reported in Figure~\ref{fig:RLDA}. The most prominent feature is of course the correct recovery of the large-${\mu}$ asymptotic behavior for both the srLDAot and srRLDAot exchange functionals. It turns out the srLDAot and srRLDAot exchange functionals give very similar exchange energies for all values of $\mu$. This comes from the fact that when going from srLDA to srLDAot Eq.~(\ref{nxHEG0neff}) tends to make the LDA exchange hole shallower and when going from srRLDA to srRLDAot Eq.~(\ref{nxRHEG0neff}) tends to make the relativistic LDA exchange hole deeper, making finally for very close descriptions. The absolute relative percentage errors of the srLDAot and srRLDAot exchange functionals are always below 10\%, and below about 2\% for ${\mu}/k_{\text{F}_{\text{max}}} \geq 0.5$.

\section{Short-range exchange generalized-gradient approximations}
\label{sec:pbe}

\begin{figure}[t]
        \centering
        \includegraphics[width=5cm,angle=270]{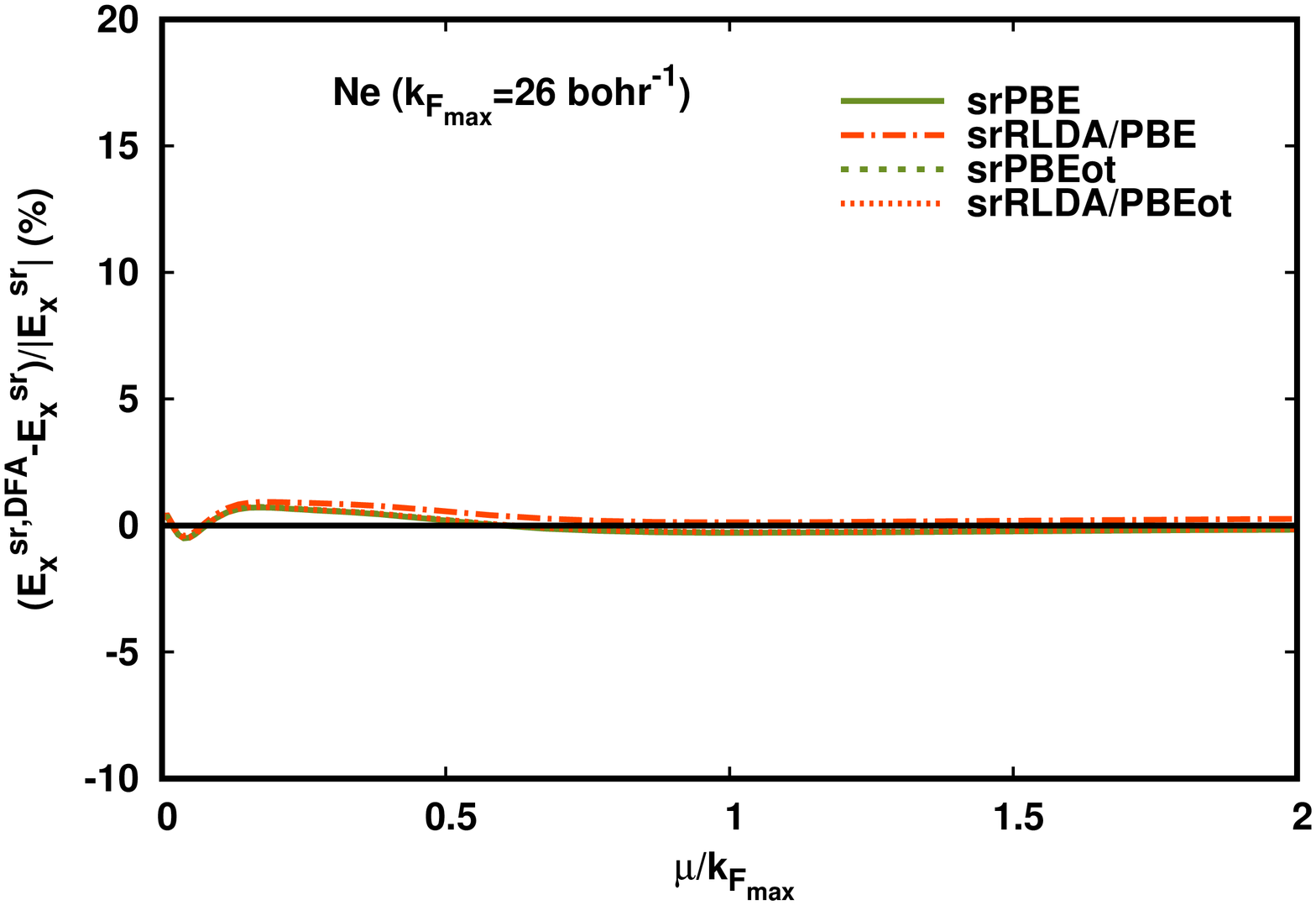}
        \includegraphics[width=5cm,angle=270]{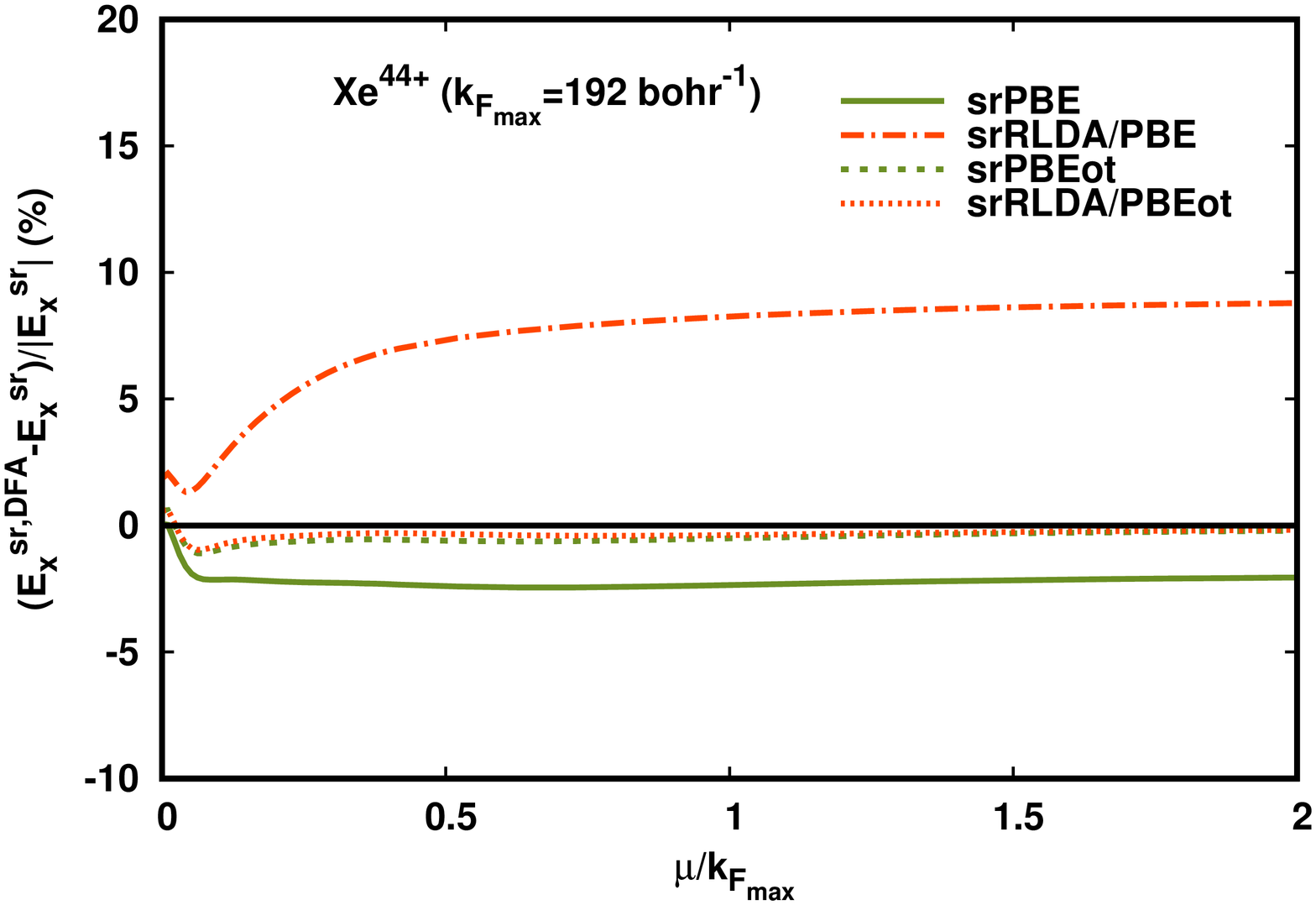}
        \includegraphics[width=5cm,angle=270]{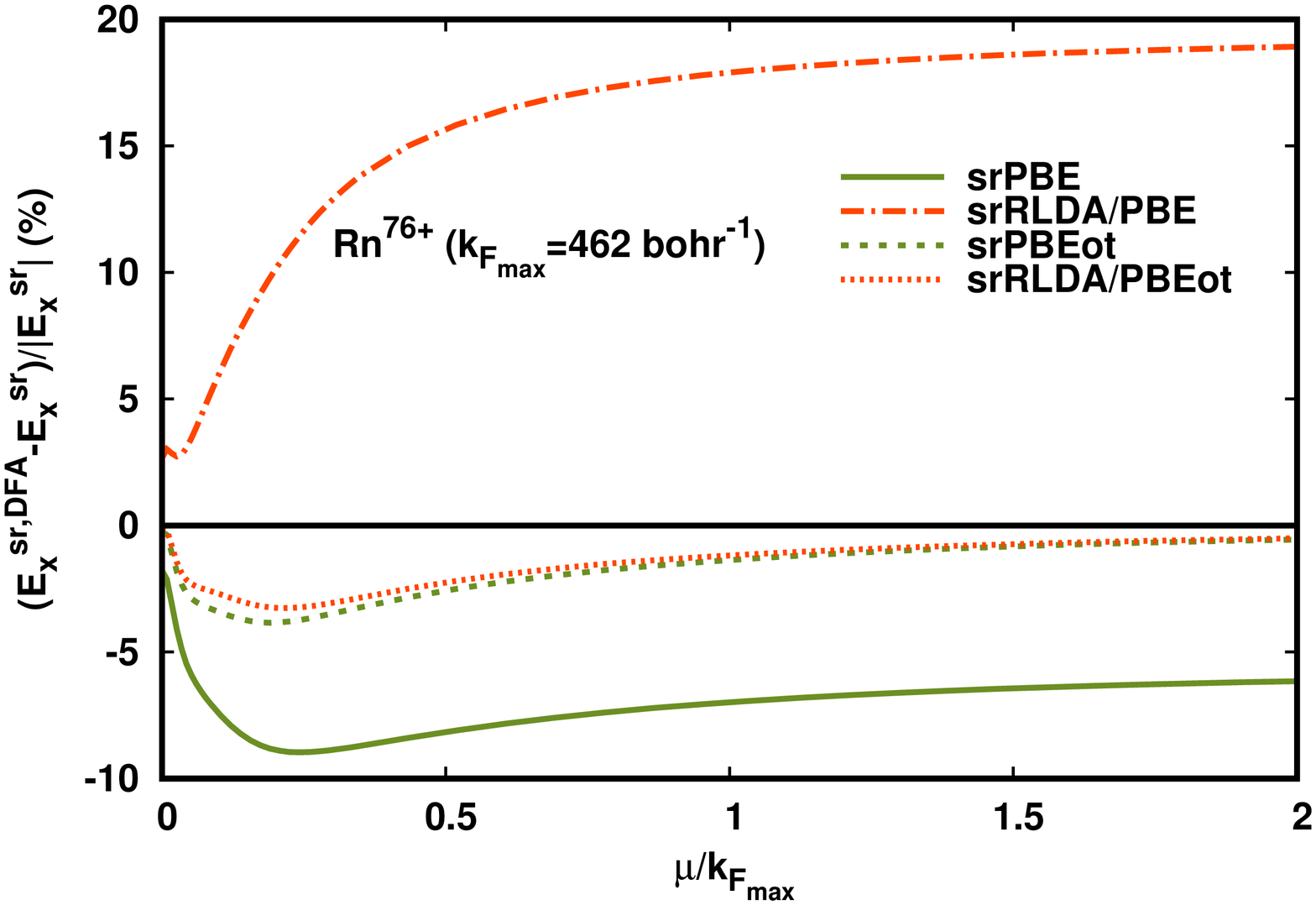}
\caption{Relative percentage error on the short-range exchange energy calculated with the srPBE, srRLDA/PBE, srPBEot, srRLDA/PBEot functionals for three representative members of the neon isoelectronic series (Ne, Xe$^{44+}$, and Rn$^{76+}$).}
\label{fig:PBE}
\end{figure}

In order to improve over the short-range LDA exchange functionals at small values of the range-separation parameter $\mu$, we now consider short-range GGA exchange functionals. We start with the non-relativistic short-range extension of the Perdew-Burke-Ernzerhof (PBE)~\cite{PerBurErn-PRL-96} of Refs.~\onlinecite{GolWerSto-PCCP-05,GolWerStoLeiGorSav-CP-06}, referred to as srPBE,
\begin{eqnarray}
E_{\text{x}}^{\text{sr,PBE},\,{\mu}}[n] &=& \int n(\b{r})~{\epsilon}_{\text{x}}^{\text{sr,HEG},\,{\mu}}\left(n(\b{r})\right) \left[1 + f_{\text{x}}^\mu (n(\b{r}),\nabla n(\b{r})) \right] \d\b{r},
\nonumber\\
\end{eqnarray}
with the function
\begin{eqnarray}
f_{\text{x}}^\mu(n,\nabla n) = \kappa - \frac{\kappa}{1+b(\mut) s^2/\kappa},
\end{eqnarray}
where $s=|\nabla n|/(2 k_\text{F}n)$ is the reduced density gradient and $\mut=\mu/(2k_F)$ is a dimensionless range-separation parameter. In this expression, $\kappa=0.840$ is a constant fixed by imposing the Lieb-Oxford bound (for $\mu=0$) and $b(\mut)=b^\text{PBE} [b^\text{T}(\mut)/b^\text{T}(0)] e^{-\alpha_\text{x} \mut^2}$ where $b^\text{PBE}=0.21951$ is the second-order gradient-expansion coefficient of the standard PBE exchange functional, $b^\text{T}(\mut)$ is a function coming from the second-order gradient-expansion approximation (GEA) of the short-range exchange energy and given in Refs.~\onlinecite{Tou-THESIS-05,TouColSav-JCP-05}, and $\alpha_\text{x}=19.0$ is a damping parameter optimized on the He atom. For $\mu=0$, this srPBE exchange functional reduces to the standard PBE exchange functional~\cite{PerBurErn-PRL-96}, and for large $\mu$ it reduces to the srLDA exchange functional.

A simple relativistic extension of this srPBE exchange functional can be obtained by replacing the srLDA part by the srRLDA one while using the same density-gradient correction $f_{\text{x}}^\mu(n,\nabla n)$, to which will refer as srRLDA/PBE,
\begin{eqnarray}
E_{\text{x}}^{\text{sr,RLDA/PBE},\,{\mu}}[n] = \;\;\;\;\;\;\;\;\;\;\;\;\;\;\;\;\;\;\;\;\;\;\;\;\;\;\;\;\;\;\;\;\;\;\;\;\;\;\;\;\;\;\;\;\;\;\;\;\;\;\;\;
\nonumber\\
 \int n(\b{r})~{\epsilon}_{\text{x}}^{\text{sr,RHEG},\,{\mu}}(n(\b{r})) \left[1 + f_{\text{x}}^\mu (n(\b{r}),\nabla n(\b{r})) \right] \d\b{r}, \;\;\;\;
\end{eqnarray}
which reduces to the srRLDA exchange functional for large $\mu$. 

The srPBE and srRLDA/PBE exchange functionals have the same (incorrect) asymptotic expansions as the srLDA and srRLDA exchange functionals [Eqs.~(\ref{ExsrLDAlargemu}) and~(\ref{ExsrRLDAlargemu})], and we can thus use the same effective densities in Eqs.~(\ref{nxHEG0neff}) and (\ref{nxRHEG0neff}) to restore their large-$\mu$ behaviors, which defines the srPBEot and srRLDA/PBEot exchange functionals
\begin{eqnarray}
E_{\text{x}}^{\text{sr,PBEot},\,{\mu}}[n] =\;\;\;\;\;\;\;\;\;\;\;\;\;\;\;\;\;\;\;\;\;\;\;\;\;\;\;\;\;\;\;\;\;\;\;\;\;\;\;\;\;\;\;\;\;\;\;\;\;\;\;\;
\nonumber\\
 \int n(\b{r})~{\epsilon}_{\text{x}}^{\text{sr,HEG},\,{\mu}}(n_\text{eff}(\b{r})) \left[1 + f_{\text{x}}^\mu (n_\text{eff}(\b{r}),\nabla n_\text{eff}(\b{r})) \right] \d\b{r}, \;\;\;\;
\end{eqnarray}
where $\nabla n_\text{eff}(\b{r}) = -2 \nabla n_\text{x}(\b{r},\b{r})$, and
\begin{eqnarray}
E_{\text{x}}^{\text{sr,RLDA/PBEot},\,{\mu}}[n] =\;\;\;\;\;\;\;\;\;\;\;\;\;\;\;\;\;\;\;\;\;\;\;\;\;\;\;\;\;\;\;\;\;\;\;\;\;\;\;\;\;\;\;\;\;\;\;\;\;\;\;\;
\nonumber\\
 \int n(\b{r})~{\epsilon}_{\text{x}}^{\text{sr,RHEG},\,{\mu}}(n_\text{eff}^\text{R}(\b{r})) \left[1 + f_{\text{x}}^\mu (n_\text{eff}^\text{R}(\b{r}),\nabla n_\text{eff}^\text{R}(\b{r})) \right] \d\b{r}, \;\;\;\;
\end{eqnarray}
where $\nabla n_\text{eff}^\text{R}(\b{r}) = [\d n_{\text{x}}^{\text{RHEG},0}(n_\text{eff}^\text{R}(\b{r}))/\d n_\text{eff}^\text{R}]^{-1} \nabla n_\text{x}(\b{r},\b{r})$.

In Figure~\ref{fig:PBE}, we report the relative percentage errors of the srPBE, srRLDA/PBE, srPBEot, and srRLDA/PBEot exchange energies for Ne, Xe$^{44+}$, and Rn$^{76+}$. For Ne, where the relativistic effects are very small, all these functionals give almost the same exchange energy, as expected. For Xe$^{44+}$ and Rn$^{76+}$, even though the srPBE and srRLDA/PBE exchange functionals are more accurate than the srLDA and srRLDA exchange functionals at $\mu=0$ (see Figure~\ref{fig:RLDA}), they eventually suffer from the same large inaccuracy as srLDA and srRLDA as $\mu$ increases. This problem is solved by using the effective densities from the on-top exchange pair density, the srPBEot and srRLDA/PBEot exchange functionals giving vanishing errors at large $\mu$. Similarly to what was observed for srLDAot and srRLDAot, the srPBEot and srRLDA/PBEot functionals give very close exchange energies for all values of $\mu$. Interestingly, we see that using the effective densities also reduces the errors of srPBE and srRLDA/PBE at $\mu=0$, making srPBEot and srRLDA/PBEot quite accurate in this full-range limit. Thus, the srPBEot and srRLDA/PBEot exchange functionals are definitely an improvement over srLDAot and srRLDAot. We observe a maximal absolute percentage error of about 3\% for Rn$^{76+}$ for $\mu /k_{\text{F}_\text{max}} \approx 0.2$.

In order to further reduce the errors, in particular for intermediate values of $\mu$, we now consider a relativistic correction to the density-gradient term in the srRLDA/PBEot exchange functional. We define a short-range relativistic PBE exchange functional using the on-top exchange pair density, referred to as srRPBEot,
\begin{eqnarray}
E_{\text{x}}^{\text{sr,RPBEot},\,{\mu}}[n] = \int n(\b{r})~{\epsilon}_{\text{x}}^{\text{sr,RHEG},\,{\mu}}(n_\text{eff}^\text{R}(\b{r})) \;\;\;\;\;\;\;\;\;\;\;\;\;\;\;\;\;\;
\nonumber\\
\;\;\;\;\times \left[1 + f_{\text{x}}^\mu (n_\text{eff}^\text{R}(\b{r}),\nabla n_\text{eff}^\text{R}(\b{r})) \; {\phi}^{\mu} (n_\text{eff}^\text{R}(\b{r})) \right] \d\b{r}, \;\;\;\;
\end{eqnarray}
where, in the spirit of the work of Engel \textit{et al.}~\cite{EngKelDre-PRA-96}, we have introduced a multiplicative relativistic correction ${\phi}^{\mu}(n)$ to the term $f_{\text{x}}^\mu (n,\nabla n)$ of the form
\begin{eqnarray}
{\phi}^{\mu} (n) &=& \frac{1 + \frac{a_{1}(\mu/c)}{\ct^{2}} + \frac{a_{2}(\mu/c)}{\ct^{4}}}{1 + \frac{b_{1}(\mu/c)}{\ct^{2}} + \frac{b_{2}(\mu/c)}{\ct^{4}}}. 
\label{phimu}
\end{eqnarray}
Since ${\phi}^{\mu} (n)$ only depends on the dimensionless parameters $\ct$ and $\mu/c$, it does not change the uniform coordinate scaling of the functional which still fulfills the scaling relation of Eq.~(\ref{Exsrgamma}).

\begin{figure}[t]
        \centering
        \includegraphics[width=5cm,angle=270]{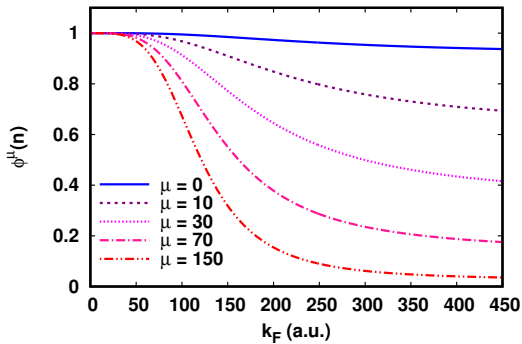}
\caption{Relativistic correction factor ${\phi}^\mu(n)$ to the density-gradient term [Eq.~(\ref{phimu})] as a function of $k_{\text{F}}$ for several values of ${\mu}$.}
\label{fig:phi}
\end{figure}

After some tests, we chose to impose $a_{1}({\mu}/c) = b_{1}({\mu}/c)$ to avoid overcorrections in low-density regions which have very small relativistic effects. We started to determine the coefficients for $\mu=0$ by minimizing the mean squared relative percentage error of the exchange energy with respect to the reference DHF exchange energy for 7 systems of the neon isoelectronic series (Ne, $\text{Ar}^{8+}$, $\text{Kr}^{26+}$, $\text{Xe}^{44+}$, $\text{Yb}^{60+}$, $\text{Rn}^{76+}$, $\text{U}^{82+}$), giving $a_{1}(0) = b_{1}(0)= 1.3824$, $a_{2}(0) = 0.3753$, and $b_{2} = 0.4096$. The resulting relativistic correction factor ${\phi}^{\mu=0} (n)$ can be seen in Figure~\ref{fig:phi}. It correctly tends to 1 in the low-density ($k_\text{F} \to 0$) or non-relativistic ($c\to \infty$) limit, and remains very close to 1 for $k_{\text{F}} \ll c$. In regions with very high densities, the relativistic correction factor ${\phi}^{\mu=0} (n)$ induces a slight reduction of the effective density-gradient correction term in the functional, reducing a bit the relative error on the exchange energy for the heaviest systems.

\begin{figure*}[t]
        \centering
        \includegraphics[width=5cm,angle=270]{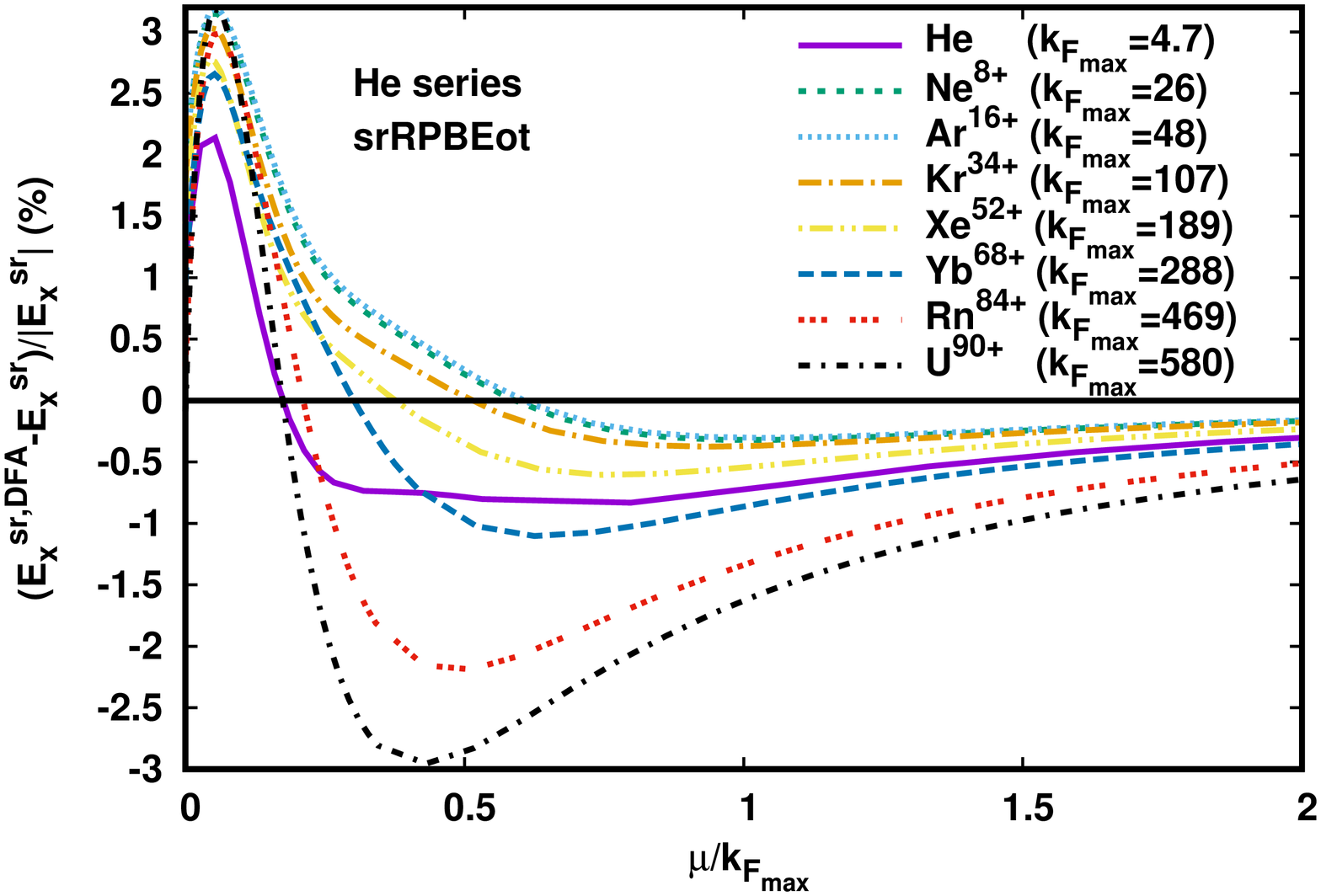}
        \includegraphics[width=5cm,angle=270]{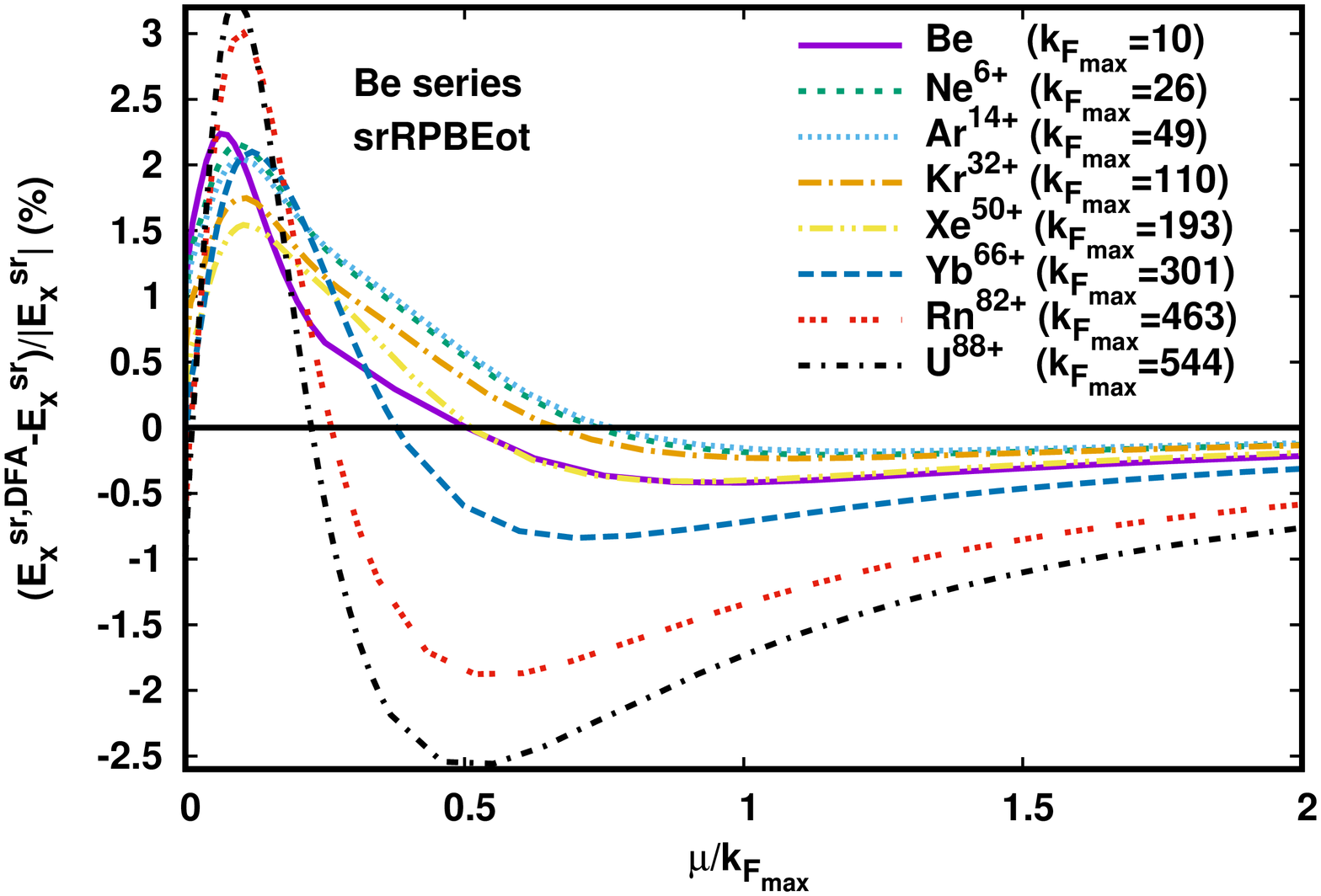}
        \includegraphics[width=5cm,angle=270]{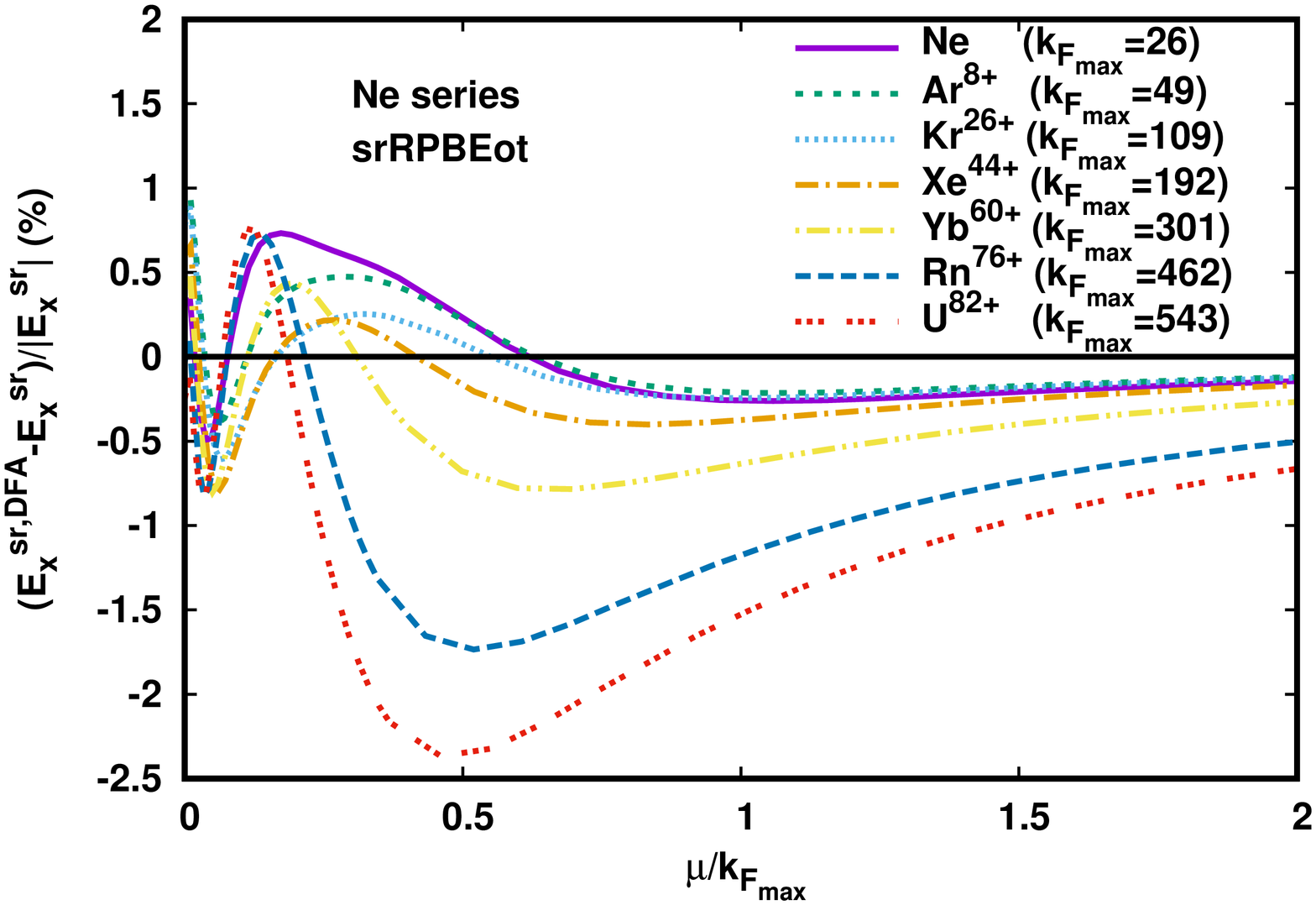}
        \includegraphics[width=5cm,angle=270]{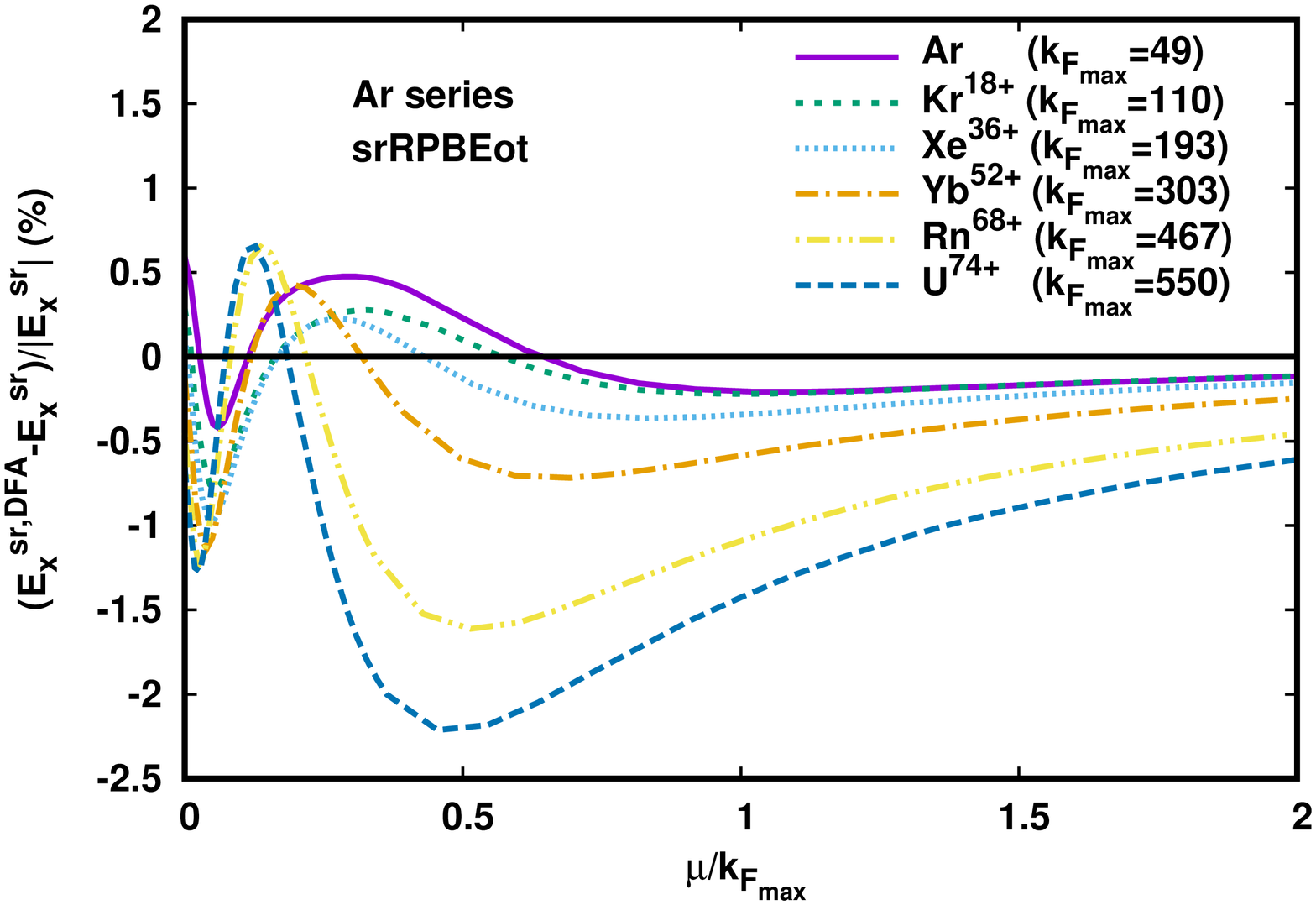}
\caption{Relative percentage error on the short-range exchange energy calculated with the srRPBEot functional for systems of helium, beryllium, neon, and argon isoelectronic series.}
\label{fig:RGGA}
\end{figure*}

For $\mu \not= 0$, we have searched for coefficients in Eq.~(\ref{phimu}) which reduce the largest errors of the srRLDA/PBEot exchange energy observed at intermediate values of $\mu$ (see Figure~\ref{fig:PBE}). We chose coefficients depending on $\mu/c$ of the form
\begin{eqnarray}
a_1(\mu/c)= b_{1}({\mu}/c) = a_1(0) [1-\text{erf}(\mu/c)],
\end{eqnarray}
\begin{eqnarray}
a_2(\mu/c)=  a_2(0) [1-\text{erf}(\mu/c)],
\end{eqnarray}
\begin{eqnarray}
b_2(\mu/c)= b_2(0) [1-{\beta}\;\text{erf}(\mu/c)],
\end{eqnarray}
with ${\beta} = - 4.235$ which has been found by minimizing the mean squared relative percentage error of the short-range exchange energy for the same 7 systems of the neon isoelectronic series and for 4 intermediate values of the range-separation parameter (${\mu}/k_{\text{F}_{\text{max}}} = 0.05; 0.1; 0.2; 0.4$). The resulting relativistic correction factor ${\phi}^{\mu} (n)$ is reported in Figure~\ref{fig:phi}. It still tends to $1$ in the low-density limit, but goes down to $0$ when ${\mu} \gg c$ in the high-density limit. The higher the value of ${\mu}$ the faster it decreases as a function of $k_{\text{F}}$.

In Figure~\ref{fig:RGGA}, we report the relative percentage errors of the srRPBEot exchange functional for systems of the helium, beryllium, neon, and argon series. For $\mu=0$, this functional achieves an error of at most about 1\% for all systems, and it has the correct large-$\mu$ limit. The maximum absolute percentage errors, which are found for intermediate values of $\mu$, tend to grow with $Z$ but remain at most about 3\% for the heavier systems. The srRPBEot exchange functional represents a significant improvement over the srPBEot and srRLDA/PBEot exchange functionals for the heavier systems.

\section{Conclusions}
\label{sec:conclusions}

In this work, we have tested the srRLDA exchange functional developed in Ref.~\onlinecite{PaqTou-JCP-18} on three systems of the neon isoelectronic series (Ne, Xe$^{44+}$, and Rn$^{76+}$) and compared it to the usual non-relativistic srLDA exchange functional. Both functionals are quite inaccurate for relativistic systems and do not have the correct asymptotic behavior for large range-separation parameter $\mu$.  In order to fix this large-$\mu$ behavior, we have then defined the srLDAot and srRLDAot exchange functionals by introducing the exact on-top exchange pair density as a new variable. These functionals recover the correct asymptotic behavior for large $\mu$ but remain inaccurate for small values of $\mu$. To improve the accuracy for small values of $\mu$, we have then developed a relativistic short-range GGA exchange functional also using the on-top exchange pair density as an extension of the non-relativistic srPBE exchange functional. Tests on the systems of the isoelectronic series of He, Be, Ne, and Ar up to $Z=92$ show that this srRPBEot exchange functional gives a maximal relative percentage error of $3\%$ for intermediate values of $\mu$ and less than $1\%$ relative error for ${\mu}=0$. Of course, in the non-relativistic limit ($c \to \infty$), all the relativistic functionals introduced in this work properly reduce to their non-relativistic counterparts.

Possible continuations of this work includes further tests on atoms and molecules, extension to the Gaunt or Breit electron-electron interactions, development of the short-range relativistic correlation functionals, and use of a local range-separation parameter.

\section*{SUPPLEMENTARY MATERIAL}
See Supplementary Material for the parameters of the even-tempered basis sets constructed in this work.

\section*{Data Availability Statement}
The data that support the findings of this study are available from the corresponding author upon reasonable request.

\appendix
\section{Uniform coordinate scaling relation for the relativistic no-pair short-range exchange density functional}
\label{app:scaling}

Here, we generalize the uniform coordinate scaling relation of the non-relativistic exchange density functional~\cite{LevPer-PRA-85} and of the non-relativistic short-range exchange density functional~\cite{TouGorSav-IJQC-06} to the case of the relativistic no-pair short-range exchange density functional of Eq.~(\ref{Exsrmu}). Since the scaling relation involves scaling the speed of light $c$, we will explicitly indicate in this section the dependence on $c$.

First, we introduce the non-interacting Dirac kinetic + rest mass energy density functional $T_\text{s}^c[n]$ defined by Eq.~(\ref{Flr}) in the special case of a vanishing range-separation parameter, $\mu=0$,
\begin{eqnarray}
T_\text{s}^c[n]  = \minmax_{\Phi_+\to n} \; \bra{\Phi_+} \hat{T}_\text{D}^c \ket{\Phi_+}
= \bra{\Phi_+^c[n]} \hat{T}_\text{D}^c \ket{\Phi_+^c[n]},
\label{Ts}
\end{eqnarray}
where $\Phi_+^c[n]$ is the relativistic Kohn-Sham single-determinant wave function. Let us now consider the scaled wave function $\Phi_{+,\gamma}^c[n]$ defined by, for $N$ electrons,
\begin{eqnarray}
\Phi_{+,\gamma}^c[n](\b{r}_1,...,\b{r}_N) = \gamma^{3N/2} \Phi_{+}^c[n](\gamma\b{r}_1,...,\gamma\b{r}_N),
\end{eqnarray}
where $\gamma>0$ is a scaling factor. The wave function $\Phi_{+,\gamma}^c[n]$ yields the scaled density $n_\gamma(\b{r}) = \gamma^3 n(\gamma \b{r})$ and is the minmax optimal wave function of $\bra{\Phi_+} \hat{T}_\text{D}^{c \gamma} \ket{\Phi_{+}}$ since it can be checked that
\begin{eqnarray}
\bra{\Phi_{+,\gamma}^c[n]} \hat{T}_\text{D}^{c \gamma} \ket{\Phi_{+,\gamma}^c[n]} = \gamma^2 \bra{\Phi_{+}^c[n]} \hat{T}_\text{D}^{c} \ket{\Phi_{+}^c[n]},
\label{}
\end{eqnarray}
and the right-hand side is minmax optimal by definition of $\Phi_{+}^c[n]$. Therefore, we conclude that
\begin{eqnarray}
\Phi_{+,\gamma}^c[n] = \Phi_{+}^{c\gamma}[n_\gamma].
\label{}
\end{eqnarray}
From the definition of the relativistic short-range exchange energy density functional $E_{\text{x}}^{\text{sr},{\mu},c}[n] = \bra{\Phi_+^c[n]} \; \hat{W}_\text{ee}^{\text{sr},{\mu}} \; \ket{\Phi_+^c[n]} - E_{\text{H}}^{\text{sr},{\mu}}[n]$, we then arrive at the scaling relation
\begin{eqnarray}
E_\text{x}^{\text{sr}, \mu \gamma, c \gamma}[n_\gamma] = \gamma E_\text{x}^{\text{sr},\mu, c}[n],
\label{}
\end{eqnarray}
or, equivalently,
\begin{eqnarray}
E_\text{x}^{\text{sr}, \mu, c}[n_\gamma] = \gamma E_\text{x}^{\text{sr},\mu/\gamma, c/\gamma}[n].
\label{Exsrgamma}
\end{eqnarray}
This scaling relation is an important constraint which is satisfied by our approximate density functionals. Besides, it shows that the low-density limit ($\gamma \to 0$) corresponds to the non-relativistic limit ($c\to\infty$), while the high-density limit ($\gamma \to \infty$) corresponds to the ultra-relativistic limit ($c\to 0$). It also shows that, for a fixed value of the range-separation parameter $\mu$, low-density regions explore the functional in the short-range limit ($\mu\to\infty$) and high-density regions explore the functional in the full-range limit ($\mu=0$). 

\section{On-top exchange pair-density in a four-component relativistic framework}
\label{app:ontoprelat}
Using four-component-spinor orbitals
\begin{eqnarray}
        \psi_i(\b{r}) &=& \left(\begin{array}{c} 
                {\psi}_i^{\text{L}{\alpha}}(\b{r})\\
                {\psi}_i^{\text{L}{\beta}}(\b{r})\\
                {\psi}_i^{\text{S}{\alpha}}(\b{r})\\
                {\psi}_i^{\text{S}{\beta}}(\b{r})\\
        \end{array}\right),
\end{eqnarray}
the on-top value of the $4\times4$ one-electron density matrix has the expression
\begin{widetext}
\begin{eqnarray}
\gamma(\b{r},\b{r}) = \sum_{i=1}^N {\psi}_{i}(\b{r})~{\psi}_{i}^{\dagger}(\b{r}) &=& 
\left(\begin{array}{cccc} 
        {\psi}_{i}^{\text{L}{\alpha}}(\b{r}){\psi}_{i}^{\text{L}{\alpha}}(\b{r})^* & {\psi}_{i}^{\text{L}{\alpha}}(\b{r}){\psi}_{i}^{\text{L}{\beta}}(\b{r})^* & {\psi}_{i}^{\text{L}{\alpha}}(\b{r}){\psi}_{i}^{\text{S}{\alpha}}(\b{r})^* & {\psi}_{i}^{\text{L}{\alpha}}(\b{r}){\psi}_{i}^{\text{S}{\beta}}(\b{r})^*\\
        {\psi}_{i}^{\text{L}{\beta}}(\b{r}){\psi}_{i}^{\text{L}{\alpha}}(\b{r})^* & {\psi}_{i}^{\text{L}{\beta}}(\b{r}){\psi}_{i}^{\text{L}{\beta}}(\b{r})^* & {\psi}_{i}^{\text{L}{\beta}}(\b{r}){\psi}_{i}^{\text{S}{\alpha}}(\b{r})^* & {\psi}_{i}^{\text{L}{\beta}}(\b{r}){\psi}_{i}^{\text{S}{\beta}}(\b{r})^*\\
        {\psi}_{i}^{\text{S}{\alpha}}(\b{r}){\psi}_{i}^{\text{L}{\alpha}}(\b{r})^* & {\psi}_{i}^{\text{S}{\alpha}}(\b{r}){\psi}_{i}^{\text{L}{\beta}}(\b{r})^* & {\psi}_{i}^{\text{S}{\alpha}}(\b{r}){\psi}_{i}^{\text{S}{\alpha}}(\b{r})^* & {\psi}_{i}^{\text{S}{\alpha}}(\b{r}){\psi}_{i}^{\text{S}{\beta}}(\b{r})^*\\
        {\psi}_{i}^{\text{S}{\beta}}(\b{r}){\psi}_{i}^{\text{L}{\alpha}}(\b{r})^* & {\psi}_{i}^{\text{S}{\beta}}(\b{r}){\psi}_{i}^{\text{L}{\beta}}(\b{r})^* & {\psi}_{i}^{\text{S}{\beta}}(\b{r}){\psi}_{i}^{\text{S}{\alpha}}(\b{r})^* & {\psi}_{i}^{\text{S}{\beta}}(\b{r}){\psi}_{i}^{\text{S}{\beta}}(\b{r})^*\\
\end{array}\right),
\end{eqnarray}
which leads to the density 
\begin{eqnarray}
n(\b{r}) = \Tr[\gamma(\b{r},\b{r})] = \sum_{i=1}^N |{\psi}_{i}^{\text{L}{\alpha}}(\b{r})|^2 + |{\psi}_{i}^{\text{L}{\beta}}(\b{r})|^2 + |{\psi}_{i}^{\text{S}{\alpha}}(\b{r})|^2 + |{\psi}_{i}^{\text{S}{\beta}}(\b{r})|^2.
\end{eqnarray}
The on-top exchange pair density has the expression
\begin{eqnarray}
n_{2,\text{x}}(\b{r},\b{r})=- \Tr[\gamma(\b{r},\b{r})^2] &=& - \sum_{i=1}^N\sum_{j=1}^{N} \Bigg( 
    |{\psi}_{i}^{\text{L}{\alpha}}(\b{r})|^2 |{\psi}_{j}^{\text{L}{\alpha}}(\b{r})|^2
+   |{\psi}_{i}^{\text{L}{\beta}}(\b{r})|^2 |{\psi}_{j}^{\text{L}{\beta}}(\b{r})|^2
+ 2 {\psi}_{i}^{\text{L}{\alpha}}(\b{r}){\psi}_{i}^{\text{L}{\beta}}(\b{r})^* {\psi}_{j}^{\text{L}{\beta}}(\b{r}){\psi}_{j}^{\text{L}{\alpha}}(\b{r})^*
\nonumber\\
&& + |{\psi}_{i}^{\text{S}{\alpha}}(\b{r})|^2  |{\psi}_{j}^{\text{S}{\alpha}}(\b{r})|^2 
   + |{\psi}_{i}^{\text{S}{\beta}}(\b{r})|^2  |{\psi}_{j}^{\text{S}{\beta}}(\b{r})|^2
  + 2 {\psi}_{i}^{\text{S}{\alpha}}(\b{r}){\psi}_{i}^{\text{S}{\beta}}(\b{r})^* {\psi}_{j}^{\text{S}{\beta}}(\b{r}){\psi}_{j}^{\text{S}{\alpha}}(\b{r})^*
\nonumber\\
&& 
+ 2 {\psi}_{i}^{\text{L}{\alpha}}(\b{r}){\psi}_{i}^{\text{S}{\alpha}}(\b{r})^* {\psi}_{j}^{\text{S}{\alpha}}(\b{r}){\psi}_{j}^{\text{L}{\alpha}}(\b{r})^* 
+ 2 {\psi}_{i}^{\text{L}{\beta}}(\b{r}){\psi}_{i}^{\text{S}{\beta}}(\b{r})^* {\psi}_{j}^{\text{S}{\beta}}(\b{r}){\psi}_{j}^{\text{L}{\beta}}(\b{r})^* \\
\nonumber
&& + 2 {\psi}_{i}^{\text{L}{\alpha}}(\b{r}){\psi}_{i}^{\text{S}{\beta}}(\b{r})^*  {\psi}_{j}^{\text{S}{\beta}}(\b{r}) {\psi}_{j}^{\text{L}{\alpha}}(\b{r})^*
   + 2 {\psi}_{i}^{\text{L}{\beta}}(\b{r}){\psi}_{i}^{\text{S}{\alpha}}(\b{r})^*  {\psi}_{j}^{\text{S}{\alpha}}(\b{r}) {\psi}_{j}^{\text{L}{\beta}}(\b{r})^*
\Bigg).
\label{n2xrrR}
\end{eqnarray}
\end{widetext}
In the non-relativistic limit, each orbital has a definite spin state, i.e. $\psi_i(\b{r})=({\psi}_i^{\text{L}{\alpha}}(\b{r}),0,0,0)$ or $\psi_i(\b{r})=(0,{\psi}_i^{\text{L}{\beta}}(\b{r}),0,0)$, and we recover the well-known expression of the on-top exchange pair density in terms of the spin densities
\begin{eqnarray}
n_{2,\text{x}}^{\text{NR}}(\b{r},\b{r}) &=& - \sum_{i=1}^N\sum_{j=1}^{N} \Bigg( 
    |{\psi}_{i}^{\text{L}{\alpha}}(\b{r})|^2 |{\psi}_{j}^{\text{L}{\alpha}}(\b{r})|^2
+   |{\psi}_{i}^{\text{L}{\beta}}(\b{r})|^2 |{\psi}_{j}^{\text{L}{\beta}}(\b{r})|^2
\Bigg) 
\nonumber\\
&=& -n_\alpha(\b{r})^2 -n_\beta(\b{r})^2,
\end{eqnarray}
or, for closed-shell systems, $n_{2,\text{x}}^{\text{NR}}(\b{r},\b{r})=  -n(\b{r})^2/2$. However, in the relativistic case, $n_{2,\text{x}}(\b{r},\b{r})$ can no longer be generally expressed explicitly with the density, as seen from the presence of terms mixing different spinor components in Eq.~(\ref{n2xrrR}). There are however two exceptions. The first exception is provided by one-electron systems for which it is easy to check that $n_{2,\text{x}}(\b{r},\b{r}) = -n(\b{r})^2$, as in the non-relativistic case. The second exception is provided by systems of two electrons in an unique Kramers pair, for which $n_{2,\text{x}}(\b{r},\b{r}) = -n(\b{r})^2/2$, as in the non-relativistic case. Indeed, for closed-shell systems, the one-electron density matrix can be decomposed into Kramers contributions
\begin{eqnarray}
\gamma(\b{r},\b{r}) = \gamma_+(\b{r},\b{r}) + \gamma_-(\b{r},\b{r}),
\end{eqnarray}
where $\gamma_+(\b{r},\b{r}) = \sum_{i=1}^{N/2} {\psi}_{i}(\b{r})\,{\psi}_{i}^{\dagger}(\b{r})$ and $\gamma_-(\b{r},\b{r}) = \sum_{i=1}^{N/2} {\bar{\psi}}_{i}(\b{r})~{\bar{\psi}}_{i}^{\dagger}(\b{r})$, and ${\bar{\psi}}_{i}(\b{r})$ is the Kramers partner of ${\psi}_{i}(\b{r})$
\begin{eqnarray}
\bar{\psi}_i(\b{r}) = \left(\begin{array}{c} 
                -{\psi}_i^{\text{L}{\beta}}(\b{r})^*\\
                {\psi}_i^{\text{L}{\alpha}}(\b{r})^*\\
                -{\psi}_i^{\text{S}{\beta}}(\b{r})^*\\
                {\psi}_i^{\text{S}{\alpha}}(\b{r})^*\\
\end{array}\right).
\end{eqnarray}
In this case, the density can then be expressed as $n(\b{r}) = 2 \Tr[\gamma_+(\b{r},\b{r})]$, and the on-top exchange pair density as
\begin{eqnarray}
n_{2,\text{x}}(\b{r},\b{r})= - 2\left( \Tr[\gamma_+(\b{r},\b{r})^2] + \Tr[\gamma_+(\b{r},\b{r})\gamma_-(\b{r},\b{r})] \right),
\end{eqnarray}
where we have used $\Tr[\gamma_+(\b{r},\b{r})^2] = \Tr[\gamma_-(\b{r},\b{r})^2]$. For an unique Kramers pair (i.e. for $N=2$), it is easy to check that $\Tr[\gamma_+(\b{r},\b{r})^2]= (\Tr[\gamma_+(\b{r},\b{r})])^2$ and $\Tr[\gamma_+(\b{r},\b{r})\gamma_-(\b{r},\b{r})]=0$, and thus
\begin{eqnarray}
n_{2,\text{x}}(\b{r},\b{r})= - 2 (\Tr[\gamma_+(\b{r},\b{r})])^2 = -\frac{n(\b{r})^2}{2} \;\; \text{for} \;\; N=2.
\end{eqnarray}
The reason why systems with one electron or two electrons in a single Kramers pair constitute exceptions is that in these systems exchange only represents in fact a self-interaction correction, and we have $E_\text{x}^{\text{sr},\mu}[n]=-E_\text{H}^{\text{sr},\mu}[n]$ for one electron and $E_\text{x}^{\text{sr},\mu}[n]=-(1/2) E_\text{H}^{\text{sr},\mu}[n]$ for two electrons in a single Kramers pair, as for the non-relativistic theory.


%
\end{document}